\documentclass{aa}  
\usepackage[normalem]{ulem}
\usepackage{color}

\usepackage{graphicx}
\usepackage{txfonts}
\usepackage{natbib}
\bibpunct{(}{)}{;}{a}{}{,}
\begin{document}

   \title{Large Interferometer For Exoplanets (LIFE):}

   \subtitle{XI. Phase-space synthesis decomposition for planet detection and characterization}
   
   \titlerunning{Phase-space synthesis decomposition}

   \author{Taro Matsuo\inst{1}
          Felix Dannert\inst{2}
          Romain Laugier\inst{3}
          Sascha P. Quanz\inst{2}
          Andjelka B. Kova{\v c}evi{\'c} \inst{4}
          \and 
          LIFE collaboration
          }

   \institute{Department of Particle and Astrophysics, Graduate School of Science, Nagoya University Furocho, Chikusa-ku, Nagoya, Aichi 466-8601, Japan\\
              \email{matsuo@u.phys.nagoya-u.ac.jp}
         \and
             ETH Zurich, Institute for Particle Physics \& Astrophysics, Wolfgang-Pauli-Str. 27, 8093 Zurich, Switzerland
          \and
             Institute of Astronomy, KU Leuven, Celestijnenlaan 200D, 3001, Leuven, Belgium\\
          \and
            Department of astronomy, Faculty of mathematics, University of Belgrade Studentski trg 16, Belgrade 11000 Serbia\\                          
             }
             
	\titlerunning{LIFE XI:PSSD for planet detection and characterization} %shortened title
	\authorrunning{Taro Matsuo et al.}%shortened list of co-authors

   \date{Received ; accepted }

  \abstract
   {A mid-infrared nulling-space interferometer is a promising way to characterize thermal light from habitable planet candidates around Sun-like stars. However, one of the main challenges for achieving this ambitious goal is a high-precision stability of the optical path difference (OPD) and amplitude over a few days for planet detection and up to a few weeks for in-depth characterization (depending on mission parameters such as aperture size, number of apertures and total instrument throughput).}
   {Here we propose a new method called phase-space synthesis decomposition (PSSD) to shorten the stability requirement to minutes, significantly relaxing the technological challenges of the mission.}
   {Focusing on what exactly modulates the planet signal in the presence of the stellar leak and systematic error, PSSD prioritizes the modulation of the signals along the wavelength domain rather than baseline rotation. Modulation along the wavelength domain allows us to extract source positions in parallel to the baseline vector for each exposure. The sum of the one-dimensional data converts into two-dimensional information. Based on the reconstructed image, we construct a continuous equation and extract the spectra through the singular value decomposition (SVD) while efficiently separating them from a long-term systematic stellar leak.}
   {We performed numerical simulations to investigate the feasibility of PSSD for the Large Interferometer For Exoplanets (LIFE) mission concept. We confirm that multiple terrestrial planets in the habitable zone around a Sun-like star at 10 $pc$ can be detected and characterized despite high levels and long durations of systematic noise. We also find that PSSD is more robust against a sparse sampling of the array rotation compared to purely rotation-based signal extraction. Using PSSD as signal extraction method significantly relaxes the technical requirements on signal stability and further increases the feasibility of the LIFE mission.} 
   
  % conclusions heading (optional), leave it empty if necessary {}
   \maketitle
%
%________________________________________________________________

\section{Introduction}

Since the Michelson stellar interferometer was mounted on the Hooker telescope and successfully measured the diameter of Betelgeuse in 1920 \citep{Michelson+1921}, ground-based interferometry has been widely used for optical, infrared, and radio astronomy \citep{Beckers+1990, Colavita+2000, Brummelaar+2005, ALMA+2015}. An image of the sky was also reconstructed based on the Van Cittert-Zernike theorem \citep{Born+1999}. The Fourier transform of the filled U-V plane provides a two-dimensional image by rotating the baseline and changing its length. 

\cite{Bracewell+1978} introduced the concept of nulling interferometry to search for exoplanet around nearby stars by introducing a $\pi$ phase shift to one of the beams of a 2-beam interferometer. When the observed sky consists of a host star and multiple planets, this concept generates a $\sin^{2}(\frac{\pi}{\lambda}\vec{B} \cdot \vec{\theta})$ fringe pattern that can null the host star at the centre of the field-of-view and transmit light from an off-axis point source, such as a planet, where $\lambda$ is the observing wavelength, $\vec{B}$ is the baseline vector, and $\vec{\theta}$ is the position vector on the sky. Rotating the baseline of the interferometer also modulates the signal of the off-axis source as a function of time, which can be leveraged for signal extraction purposes. Followed by the proposal of mid-infrared nulling interferometry, \cite{Angel+1986} noticed that the mid-infrared wavelength range is useful for the characterization of temperate Earth-like planets. This is because of the relatively low contrast between the planet and its host star compared to that observed in the visible wavelength range. In addition, $CH_{4}$ and $O_{3}$ are atmospheric biosignatures that have strong absorption bands in the same wavelength range \citep[e.g.,][]{DesMarais+2002, Fujii+2018}. A combination of these two studies led to the construction of a concept for remotely measuring the activity of primitive life on distant planets through detecting a variety of $CH_{4}$ and $O_{3}$, named $Darwin$ \citep{Leger+1996}. Darwin was an ESA-led concept and similar parallel activities were going on on the US side in the context of Terrestrail Planet Finder-Interferometer \citep[TPF-I;][]{Lawson+2008}.

\cite{Angel+1997} used a cross-correlation technique to efficiently find the modulated signal while rotating the baseline, leveraging Bracewell's idea. A nuller consisting of four apertures was also introduced to obtain a fourth-order null of the host star. Furthermore, \cite{Mennesson+1997} proposed five collectors to suppress modulation of the exozodiacal light during baseline rotation, keeping the fourth-order null. Instead, \cite{Velusamy+2003} mentioned the advantage of a dual Bracewell interferometer consisting of two equivalent second-order nullers, which overcomes the ambiguity of the planet positions with a phase chop. We note that a phase shifter ($\pi / 2$ for nulling interferometers) is introduced to one of the two beams, and the two states are formed by inserting or removing the phase shifter. In addition, the subtraction of the two chopped states can separate symmetric components, including stellar leakage and background light, from the off-axis point sources. Finally, both the TPF-I and Darwin mission concepts favored a dual Bracewell interferometer \citep{Cockell+2009}. Although TPF-I and Darwin were anticipated to detect and characterize the thermal emissions from Earth-like planets for the first time, they were postponed indefinitely due to the technical difficulties.

However, stellar leak is more sensitive to the optical path difference (OPD) and low-order aberrations in the second-order null than in the fourth-order null \citep[e.g.,][]{Hansen+2022b}. \cite{Lay+2004} quantified the systematic errors generated from the fluctuation of the null depth, which could obscure the modulated planet signal over the baseline rotation. The ideal phase chop technique can successfully remove most of the systematic noise errors, and only the first-order phase error and the cross-term of phase and amplitude errors remain in the demodulated signal. To identify an Earth-like planet around a Sun-like star at 10 $\mu m$, the OPD and amplitude perturbations need to be stabilized to 1.5 $nm$ and 0.1\%, respectively \citep{Lay+2004}. Because observing such a planet with a signal-to-noise ratio of seven requires an integration time of a few days, according to preliminary analyses for the LIFE mission presented in \cite{Dannert+2022}, the 1.5-$nm$ OPD stability requirement holds for the same period. This imposes strict requirements on the formation flight and the optical beam transport and combination system. \cite{Lay+2006} proposed to stretch the aspect ratio of the rectangular four collector array of the Double Bracewell interferometer to remove instrumental noise induced by the systematic effects from the data to mitigate the requirements by a factor of 10. The method utilized different behaviors between the planet and instability noise achieved by stretching the ratio between the nulling and imaging baseline to 1:6. While this relaxes the requirements on the null stability, the streching of the baselines requires a more fuel consumption compared to baseline rotation. 

Instead of using the modulated signals of off-axis objects during baseline rotation, \cite{Matsuo+2011} proposed a method for estimating the positions of off-axis objects and obtaining their spectra from a few baselines, focusing on the modulation of the signal along the wavelength domain. This method requires only relative stability among the wavelength channel across the observed spectrum instead of stability of the null depth during the baseline rotation. In addition, when the number of baselines is larger than that of the detectable objects in the field of view of the interferometer, one can effectively separate planet signals from long-term fluctuations caused by systematic effects. The extended method optimizes for continuously rotating the baseline instead of fixing baselines, which could further mitigate the stability requirements of a space-based nulling interferometer. The present study is complementary to developments of the formation flying interferometry \citep[e.g.,][]{Hansen+2022, Matsuo+2022} and ground-based nulling experiments \citep[e.g.,][]{Ertel+2020, Ranganathan+2022}. These efforts provide a support for the Large Interferometer for Exoplanets (LIFE), which represents a science theme that was recognized as one out of three potential science themes for a future L-class mission in the Voyage 2050 of the European Space Agency. Based on the heritage of Darwin and TPF-I, but leveraging most recent scientific and technological developments, LIFE will directly detect and characterize the thermal light from habitable planet candidates. LIFE could detect 25 - 45 terrestrial planets in the habitable zone around nearby F-, G-, K-, and M-type stars under conservative assumptions \citep{Quanz+2022, Kammerer+2022}. 

In the following we propose the phase-space synthesis decomposition (PSSD) method for extracting the planet signal. This method could mitigate the rigorous requirements imposed on the nulling space interferometer, which is complementary to ongoing technological demonstrations for LIFE. Section 2 provides a brief overview of PSSD and its mathematical explanation. We perform a numerical simulation to investigate the feasibility of the method using the LIFE simulator \citep{Dannert+2022} in Section 3. Section 4 discusses the limitation of PSSD and its advantages and disadvantages. We conclude with our main findings in Section 5.

\section{Concept} \label{sec:concept}

This section provides an overview of PSSD and then introduces its procedures from planet detection to spectral characterization. The analytical equations constructed for PSSD explain the two processes, planet detection and characterization, in detail. 

\subsection{Overview} \label{subsec:overview}
PSSD is divided into two processes: (1) search for the planet signal, and (2) measurement of the planet spectrum. Both steps require a continuous baseline rotation and the same operation as the previous method that extracts the modulated signal through the cross-correlation or the maximum likelihood of the data obtained while rotating the baseline \citep[e.g.,][]{Angel+1997, Dannert+2022}. However, the previous method and PSSD differ in how the planetary positions are reconstructed. The previous process transforms interferometric signals collected by spinning the baseline into the planet position by fitting the modulation in both the azimuth and wavelength domains simultaneously. However, focusing on the fact that the spectrum of a G-type star is smoothly distributed at mid-infrared wavelengths \citep[e.g.,][]{Husser+2013}, a one-dimensional image in parallel to baseline is first formed by correlation of the interferometric signal only along the wavelength domain, which basically has the same characteristics as Fourier transform of the signal in the same direction \citep{Matsuo+2011}. Because the wavelength dependence of the stellar leak is largely different from that of the planet signal, the partial correlation only along the wavelength domain can decompose the stellar leak and planet signal. After spinning the baseline, summing over one-dimensional images, instead of a cross-correlation of the signal in the azimuth domain, transforms a set of one-dimensional positional information into two-dimensional positional information. Thanks to the partial correlation only in the wavelength domain, an impact of a long-term systematic error on image reconstruction can be avoided.

PSSD receives both the advantages of the two previous methods, the cross-correlation method \citep{Angel+1997} and Fourier transform of the signal along the wavelength domain \citep{Matsuo+2011}. While the corss-correlation method increases the signal-to-noise ratio as much as possible, the latter efficiently decomposes the stellar leak and planet signal. PSSD combines the two methods by employing a local cross-correlation of the signal only along the wavelength domain, instead of the full cross-correlation in both the wavelength- and time-domains. Generally, local (or segmented) cross-correlation is used when cross-correlating data in smaller segments. We can directly compare small sections of two arrays of data by cross-correlating corresponding segments, allowing for a more localised analysis. This method is very useful when analyzing complex astronomical phenomena with fluctuations or patterns in various portions of compared signals \citep[e.g.,][]{Kovacevic+2018}.

Thanks to the combination of the two previous methods, PSSD provides three advantages in terms of planet detection. First, PSSD could shorten the required stability duration from a few days to a few minutes. Second, PSSD could also mitigate the impact of the limited number of baselines on search for the planet signal. Third, PSSD could have robustness against a larger OPD fluctuation. Utilizing the advantages of the planet detection process, PSSD also develops a method for extracting the planet spectrum embedded in the stellar leak. 

Regarding the first advantage, the required stability duration could be shortened from a few days to a few minutes because we do not use the correlation of the planet signals collected while rotating the baseline. This is equivalent to the period for obtaining the two-phase chop states. We note that the period of switching between the two-phase chop states is determined such that the slow change of the background can be fully sampled \citep{Absil+2003}. PSSD only requires relative stability along the wavelength (i.e. among the spectral data), rather than the stability of signals received while turning the baseline. The continuous and wide wavelength range obtained from space, such as 4 to 18.5 $\mu m$ for the LIFE observatory, realizes this alternative approach. We note that what type of object (e.g. Jovian planet or terrestrial planet) orbits the host star is unclear in the planet detection phase because the light is integrated over the entire wavelength range (4 to 18 $\mu m$ for the LIFE mission) in this phase. Regarding the second advantage, because PSSD reconstructs a one-dimensional image from one imaging baseline, two-dimensional positionalal information can be extracted from fewer baselines. In other words, PSSD has more resistance against a limited number of data collected during baseline rotation than the previous cross-correlation technique. In terms of the last advantage, a fluctuation of OPD during baseline rotation does not correlate with a modulation of the planet signal along the wavelength domain. Instead, the OPD error contributes to the observing data as a noise because the stellar leak is inversely proportional to approximately the fourth power of wavelength. PSSD can detect the planet light unless the modulation of the planet signal is embedded in the stellar leak. Thus, PSSD is more robust against a large OPD error in terms of planet detection.

Next, the planetary spectra are derived based on the positional information of the planets. Because we calculate the modulation of planet light while rotating the baseline based on the information of the estimated planet position, the planet light for each spectral channel can be extracted by fitting the data through the singular value decomposition (SVD) method. However, because the modulation of the planet signal during baseline rotation is used for the reconstruction of the planet spectrum, the reconstructed spectrum is more affected by a long-term systematic error compared to planet detection. As a result, the large contrast between the stellar leak and a temperate planet at short wavelengths prevents us from precisely reconstructing the planet spectrum in the same wavelength range (Section \ref{subsec:limitation}). Before applying the data to the SVD method, the stellar leak has to be subtracted from the data if the stellar leak is much brighter than the planet light. On the other hand, it is difficult to measure the OPD change because the number of available photons is very limited at the nulled output. 

Here we find that the stellar leak induced by the systematic OPD error could be measured from the data at short wavelenths. Because warm and temperate planets are much fainter than the stellar leak in the short wavelength range, only the stellar leak mainly contributes to the data. Since there is no a strong chromatic aberration in the optical system, thanks to reflectors constructing the optical system of LIFE, the stellar leak may be able to be expressed as a function of wavelength. For example, when the stellar leak is induced by the OPD error, the stellar leak is inversely proportional to approximately the fourth-power of wavelength. Since there are a large volume of the data collected during baseline rotation, we could estimate the wavelength dependence of the stellar leak using a simple model, such as an exponential function. We note that modeling the wavelength dependence of the stellar leak is already performed in the data reduction pipeline of GRAVITY \citep[e.g.,][]{Nowak+2020}. Once the wavelength dependence is derived, the modeled stellar leak could be extrapolated to a longer wavelength range. The modeled stellar leak is subtracted from the data and is not, in principle, contaminated in the reconstructed spectra. Because the stellar leak is much weaker at the wavelengths longer than 10 $\mu m$, the planet spectrum is less affected even if there exists a chromatic aberration. 

We note, however, that if a Jovian planet close to its host star exists in the observing object, the planet is brighter than the stellar leak even at short wavelengths. Subtracting the bright planet from the data is required for modeling the wavelength dependence of the stellar leak at short wavelengths. Thanks to the long imaging baseline, an inner planet can be spatially resolved from the host star. Because both the position and spectrum of the inner planet are obtained through PSSD, we can estimate how the planet signal is modulated during baseline rotation and subtract it from the data.

We also need to emphasize that PSSD is validated only for objects having smoothed spectra such as a Planck function. If the spectra have sine components, the positions of the objects are shifted from the true positions. The reconstructed spectra are systematically affected by the wrongly estimated positions. Since the low-dispersion spectra of the atmospheres in exoplanets are close to a Planck function, the systematic shifts are not considered in this study. We note that the reflected light of Europa from the Sun in optical does not affect the position on the reconstructed image in spite of the spectrum including a number of lines \citep{Matsuo+2022}.

Here we overview the concrete data reduction of this method. The process of the planet detection consists of the following five steps.
\begin{itemize}
   \setlength{\leftskip}{0.5cm}
    \item[1] Subtract the two chop states for each baseline to obtain the sine component of the complex visibility.
    \item[2] Extract the modulated signals of the off-axis point sources along the wavelength (see Section \ref{subsec:planet_detection}). 
    \item[3] Repeat processes [1] and [2] during rotation of the baseline. This process is done for each set of the two-phase chop states.
    \item[4] Transform a set of reconstructed one-dimensional images into a two-dimensional image (i.e. phase-space synthesis).
    \item[5] Search for planet light in the reconstructed two-dimensional image and measure the planet position if it exists.    
\end{itemize}    
The characterization process is as follows:
\begin{itemize}
	\setlength{\leftskip}{0.5cm}
    \item[6] Perform procedure [1] with longer integration time.
    \item[7] Model the wavelength dependence of the stellar leak from the collected data at short wavelengths (e.g., 4 to 6 $\mu m$) if the stellar leak is much brighter than the planet light due to a large OPD error.
    \item[8] Subtract the stellar leak from the collected data after extrapolating the stellar leak model constrcuted in process [7] to the long wavelength range. 
    \item[9] Construct a matrix equation of the following form from the set of collected data: $O = R I$, where $O$ is the observable vector, $R$ is the response function, and $I$ is the vector of the input sky.
    \item[10] Solve the matrix equation using the SVD method to extract the planet spectrum (i.e. phase-space decomposition).
\end{itemize}    

The phase information is summed for planet detection through processes [1] to [5]. In contrast, the phase information is decomposed in the planet characterization phase through procedures [6] to [10]. If the stellar leak is not much brighter than the planet light at short wavelengths, processes [7] and [8] can be skipped. As shown in Section \ref{subsec:results}, when the systematic OPD RMS error is 0.75 $nm$, corresponding to the standard requirement of LIFE \citep{Dannert+2022}, the planet spectra can be precisely extracted without processes [7] and [8]. Although processes [1] and [6] are equal in data reduction, the required integration time is different. Because the planet light is integrated over the entire wavelength range in the reconstructed image through processes [1] – [5], the required integration time for planet detection is much shorter than that for planet spectrum obtained through processes [6] - [10].
% An integration time of a few days is sufficient for detecting a temperate terrestrial planet for the LIFE mission concept with four 2-m-class telescopes and 5 \% throughput. Achieving a signal-to-noise ratio of more than ten for characterizing planets requires a few tens of days, which is longer than that for planet detection by a factor of the resolving power. 

We explain the planet detection and characterization in Sections \ref{subsec:planet_detection} and \ref{subsec:characterization}, respectively, constructing equations for PSSD.

\subsection{Search for planet signal (phase-space synthesis)} \label{subsec:planet_detection}

When a dual-Bracewell nulling interferometer with a $\frac{\pi}{2}$ phase chop observes the sky, the observed two-chop states in the unit of photoelectrons are as follows \citep[e.g.,][]{Beichman+1999, Matsuo+2011}
\begin{eqnarray}
	\label{eqn:signal}
	O_{\pm} (\lambda) &=& \frac{1}{2} \int \int d^{2}\theta I (\lambda, \vec{\theta}) \sin^{2}\left(\frac{\pi}{\lambda} \vec{b} \cdot \vec{\theta} + \delta l_{n}\right) \\ \nonumber
	&\times & \left \{1 \pm \sin \left( \frac{2 \pi}{\lambda} \vec{B} \cdot \vec{\theta} + \delta l_{i}\right) \right \},
\end{eqnarray}
where $\vec{\theta}$ is the position vector in the sky, $\vec{b}$ and $\vec{B}$ are the nulling and imaging baseline vectors, $I(\lambda, \vec{\theta})$ is the signal without the effect of sky transmission caused by the dual-Bracewell nulling interferometer at wavelength of $\lambda$, and $\delta l_{i}$ and $\delta l_{n}$ are the optical path differences of the imaging and nulling baselines, respectively. The $(+)$ and $(-)$ notations indicate the two chop states. In Equation \ref{eqn:signal}, we assumed that the two nulling baselines for the dual-Bracewell nulling interferometer have the same optical path difference error, $\delta l_{n}$, for simplicity.

The planetary system is the sum of the host star $I_{*} (\lambda, \vec{\theta})$, multiple planets $N_{p}$, $\Sigma_{k}^{n} I_{p,k}$, local zodiacal light, $I_{lz} (\lambda)$, and the exozodiacal light $I_{ez} (\lambda, \vec{\theta})$. The spectrally resolved signal for the planetary system is written as
\begin{eqnarray}
	\label{eqn:signal_for_system}
	O_{\pm} (\lambda) &=& \frac{1}{2} \int \int_{\Omega_{*}} d^{2}\theta I_{*} (\lambda, \vec{\theta}) \sin^{2}\left(\frac{\pi}{\lambda} \vec{b} \cdot \vec{\theta}+ \delta l_{n} \right)  \\ \nonumber
	&\times & \left \{1 \pm \sin \left( \frac{2 \pi}{\lambda} \vec{B} \cdot \vec{\theta} + \delta l_{i}  \right) \right \} \\ \nonumber
&+& \frac{1}{2} \Sigma_{k}^{N_{p}} I_{p, k} \left( \lambda, \vec{\theta}_{p,k} \right) \Omega_{p,k}  \sin^{2}\left(\frac{\pi}{\lambda} \vec{b} \cdot \vec{\theta}_{p,k} + \delta l_{n} \right)  \\ \nonumber
  &\times & \left \{1 \pm \sin \left( \frac{2 \pi}{\lambda} \vec{B} \cdot \vec{\theta}_{p,k} \right) + \delta l_{i} \right \} \\ \nonumber
&+& \frac{1}{2} \int \int_{\Omega_{fov}} d^{2}\theta \left( I_{lz} (\lambda) + I_{ez} (\lambda, \vec{\theta}) \right) \sin^{2}\left(\frac{\pi}{\lambda} \vec{b} \cdot \vec{\theta} + \delta l_{n} \right)  \\ \nonumber
&\times & \left \{1 \pm \sin \left( \frac{2 \pi}{\lambda} \vec{B} \cdot \vec{\theta} + \delta l_{i}\right) \right \},
\end{eqnarray}
where $\Omega_{*}$ and $\Omega_{p,k}$ are the solid angles of the host star and the $k$-th planet, respectively, $\Omega_{fov}$ is the field of view of the interferometer, and $\theta_{p,k}$ is the position vector of the $k$-th planet. As shown above in step [1], the demolutated signal is given by
\begin{eqnarray}
	\label{eqn:demodulated_signal}
	O (\lambda) &=& O_{+}(\lambda) - O_{-} (\lambda) \\ \nonumber
	&=& \frac{1}{2} \int \int_{\Omega_{*}} d^{2}\theta I_{*} (\lambda, \vec{\theta}) \sin^{2}\left(\frac{\pi}{\lambda} \vec{b} \cdot \vec{\theta} + \delta l_{n} \right) \\ \nonumber
	&\times & \sin \left( \frac{2 \pi}{\lambda} \vec{B} \cdot \vec{\theta} + \delta l_{i} \right) \\ \nonumber
&+& \frac{1}{2} \Sigma_{k}^{N_{p}} I_{p, k} \left( \lambda, \vec{\theta}_{p,k} \right) \Omega_{p,k}  \sin^{2}\left(\frac{\pi}{\lambda} \vec{b} \cdot \vec{\theta}_{p,k} + \delta l_{n} \right) \\ \nonumber
&\times & \sin \left( \frac{2 \pi}{\lambda} \vec{B} \cdot \vec{\theta}_{p,k} + \delta l_{i} \right) \\ \nonumber
&+& \frac{1}{2} \int \int_{\Omega_{fov}} d^{2}\theta I_{ez} (\lambda, \vec{\theta}) \sin^{2}\left(\frac{\pi}{\lambda} \vec{b} \cdot \vec{\theta} + \delta l_{n}\right) \\ \nonumber
&\times & \sin \left( \frac{2 \pi}{\lambda} \vec{B} \cdot \vec{\theta} + \delta l_{i} \right),
\end{eqnarray}
where the local zodiacal light was assumed to be removed from the demodulated signal because of its symmetrical structure. If the host star is perfectly positioned at the center of the field of view (FOV), the stellar leak disappears in Eq. \ref{eqn:demodulated_signal} and contributes only as shot noise. Now we move to step [2]. 

\begin{figure}
	 \centering
	\includegraphics[scale=0.1,height=1.5cm,clip]{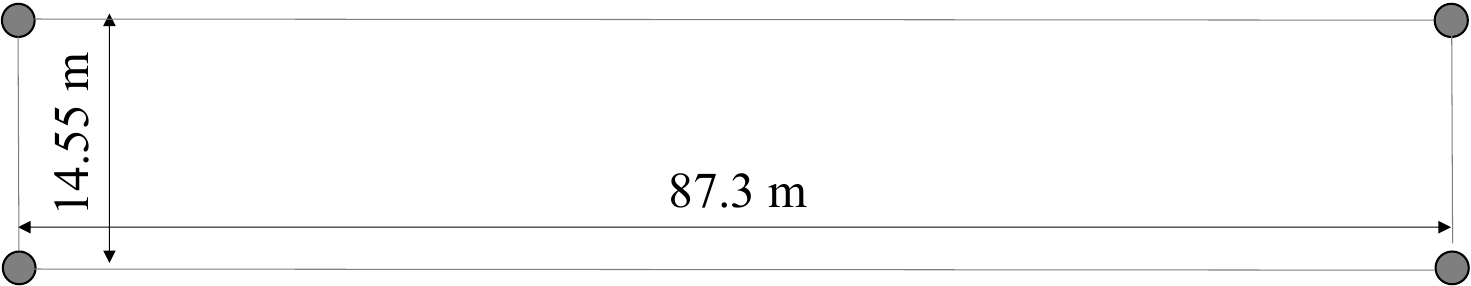}
	\caption{Rectangular array used for this study. Each filled circle represents a 2$m$-diameter telescope. The ratio of the imaging to nulling baselines is 6:1.}
	\label{fig:configuration}
\end{figure}

There are two approaches for step [2]: the cross-correlation method \citep{Angel+1997} and the Fourier transform \citep{Matsuo+2011}. While the former focuses on the modulated signal while rotating the baseline, the latter uses the modulated one along the wavelength domain for each baseline. Here, we combine the advantages of the two methods. We use the correlation method \citep{Angel+1997} to extract the signal correlated to the modulation of the planet along the wavelength and derive the planet position for each baseline. After rotation of the baseline, the positions of the planets are obtained. We employ a rectangular array configuration with a baseline ratio of 6:1 based on the baseline of the LIFE mission concept (see Figure \ref{fig:configuration}). The parameters of the configuration are the same as those used for the numerical simulations in Section \ref{sec:simulation}. The configuration is optimized to maximize the throughput of the habitable zone around a G-type star at 10 $pc$.

Given that the position vector of the correlated signal is $\vec{\alpha}_{corr}$ (i.e. the two-dimensional position of the signal), the positional information reconstructed from the $j$-th imaging baseline vector, $\vec{B}_{j}$, is
\begin{equation}
	\label{eqn:reconstructed_corr}
	M_{corr,j}(\alpha_{j}) = \Sigma_{i}^{N_{i}} O(\lambda_{i}) \sin \left(\frac{2\pi}{\lambda_{i}} \vec{B}_{j} \cdot \vec{\alpha}_{corr} \right) \sin^{2} \left(\frac{\pi}{\lambda_{i}} \vec{b}_{j} \cdot \vec{\alpha}_{corr} \right),
\end{equation}
where $\lambda_{i}$ is the $i$-th spectral element, and $N_{i}$ is the number of elements. $M_{corr,j}$ tells us about the position of the correlated signal, $\alpha_{j}$, projected to the baseline vector, $\vec{B}_{j}$. As shown in Equation \ref{eqn:demodulated_signal}, $O(\lambda_{i})$ is equal to the sum of the stellar leak, planet signals, and background components. Each component has a different spectrum energy distribution (see Figure \ref{fig:setup}(b)).

\begin{figure}
	 \centering
	\includegraphics[scale=0.1,height=12cm,clip]{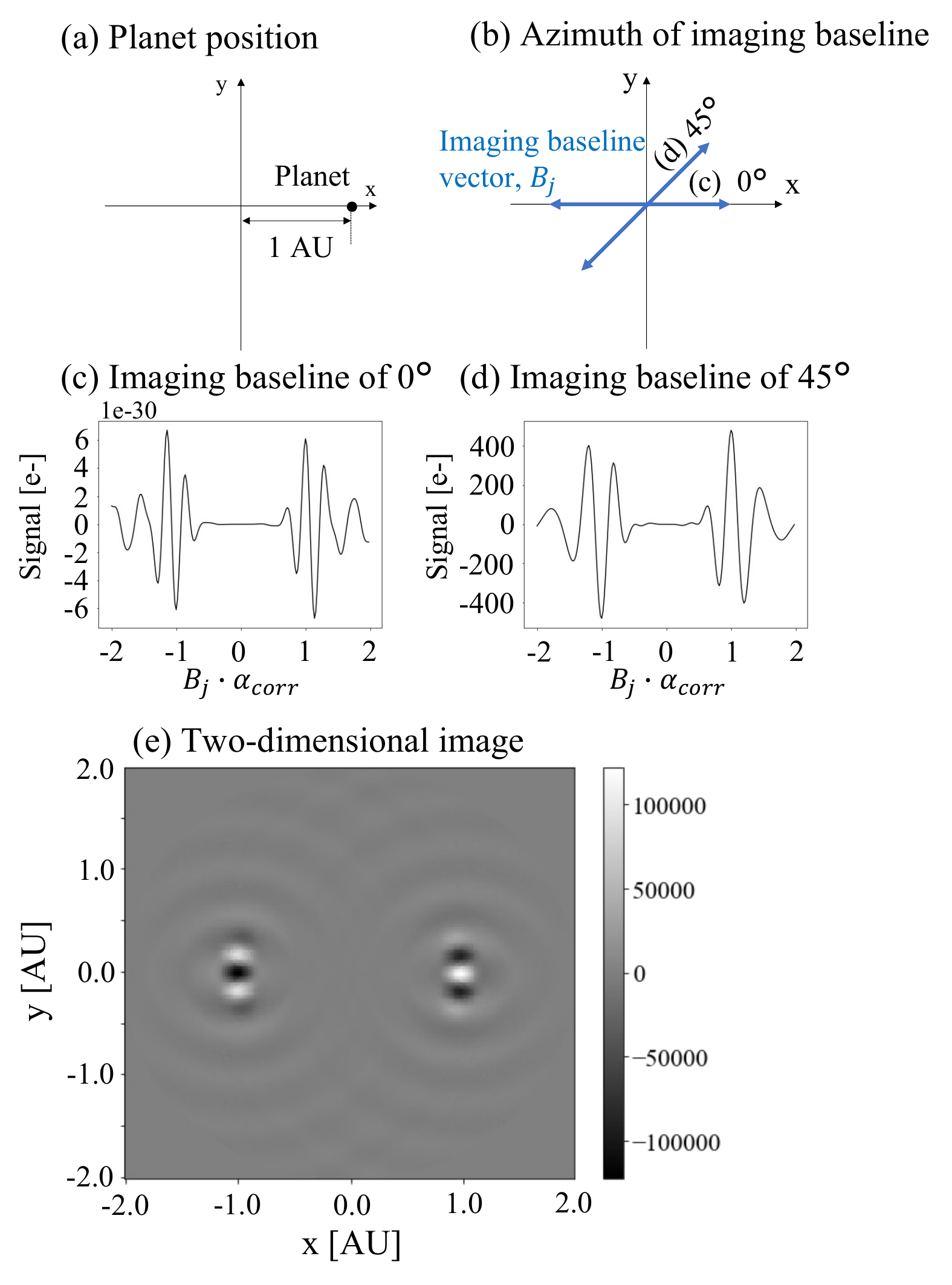}
	\caption{Procedure of image reconstruction. Panel (a) shows the planet position for this simulation. The planet is positioned at 1 $AU$ in parallel to the $x$ axis, where $(x, y)$ is the coordinate system of the object plane in the unit of $AU$.} The planetary system at 10 $pc$ consists of only one Earth-radius planet with an effective temperature of 285 $K$ (without its host star). Panel (b) shows the directions of the imaging baseline for the two reconstructed one-dimensional images (panels (c) and (d)). Panel (e) shows a two-dimensional image converted from the set of the one-dimensional images. The vertical axis of panels (c) and (d) and the color bar of panel (e) represent the number of photoelectrons. The parameters of the telescope and instrument are compiled in Table \ref{tab:instrument}.
	\label{fig:image_reconstruction}
\end{figure}

In order to explain how PSSD works, we perform a simulation under a simple condition that an Earth-like planet is positioned at 1 $AU$ parallel to the $x$-axis (panel (a) of Figure \ref{fig:image_reconstruction}). When the azimuths of the imaging baseline are 0 and 45$^{\circ}$ (panel Fig. \ref{fig:image_reconstruction} (b)), two one-dimensional images parallel to the imaging baseline vector are generated (panels (c) and (d) of Figure \ref{fig:image_reconstruction}). When the azimuth of the imaging baseline is 0$^{\circ}$, the planet light is nulled for the rectangular array because the nulling baseline is parallel to the $x$-axis (panel Fig. \ref{fig:image_reconstruction}(c)). In fact, the peak value of the planet is almost 0. In contrast, for the azimuth angle of 45$^{\circ}$, the planet light is extracted at 1 $AU$ of the $x$ axis (panel Fig. \ref{fig:image_reconstruction} (d)). The peak value is a much larger than that of the nulled planet. The data obtained by the rectangular array also has information on the planet position in the direction perpendicular to the imaging baseline, thanks to the nulling baseline. The planet position along the nulling baseline is weakly constrained for each angle of the imaging baseline. Thus, both the imaging baseline and the nulling baseline can be utilized for planet detection. 

Because the stellar leak caused by the OPD error is inversely proportional to approximately $\lambda^{-4}$, the first term of $O(\lambda_{i})$ in Equation \ref{eqn:demodulated_signal} does not correlate with $\sin \left( \frac{2 \pi}{ \lambda_{i}} \vec{B_{j}} \cdot \vec{\alpha}_{corr} \right) \sin^{2} \left( \frac{\pi}{ \lambda_{i}} \vec{b_{j}} \cdot \vec{\alpha}_{corr} \right)$ along the wavelength domain. The OPD error does not have less influence on the reconstruction of one-dimensional images, compared to the previous method that extracts the modulated signal through the cross-correlation of the data collected while rotaing the baseline. In other words, PSSD has robustness against a large OPD fluctuation, which is the third advantage of PSSD, as discussed in Section \ref{subsec:overview}.

The two-dimensional positional information is obtained by summing the one-dimensional images collected while rotating the baseline:
\begin{equation}
	\label{eqn:reconstructed_corr_final}
	M_{corr}(\vec{\alpha}_{corr}) = \Sigma_{j}^{N_{j} }M_{corr,j}(\alpha_{j}),
\end{equation}
where $N_{j}$ is the number of collected baselines. Panel (e) shows the two-dimensional image reconstructed through Equation \ref{eqn:reconstructed_corr_final}. The pixel value at the planet position on the two-dimensional image corresponds to the sum of the pixel values at the same position on the one-dimensional images. Because PSSD focuses on the signal modulated by the wavelength for each baseline instead of one modulated by the rotation of the baseline, PSSD is less affected by the long-term systematic noise than the purely rotation-based signal extraction, which is the first advantage of PSSD discussed in Section \ref{subsec:overview}. The required stability duration could be shortened to a few minutes, corresponding to the period for obtaining the two-phase chop states.

PSSD is not also less impacted by a sparse U-V sampling. We discuss how the limited number of collected baselines affects the reconstructed two-dimensional image in Section \ref{subsec:comparison}. 

We also note the relationship between the correlation and Fourier tranform methods. The Fourier transformation of the demodulated signal along the wavelength gives a one-dimensional image of the sky in parallel to the $j$-th imaging baseline vector:
\begin{equation}
	\label{eqn:ft_image}
	M_{FT,j}(\alpha_{FT,j}) = \int d\left(\frac{1}{\lambda} \right) O(\lambda) \sin\left(\frac{2\pi}{\lambda} |\vec{B}_{j}| \alpha_{FT,j} \right) \sin^{2}\left(\frac{\pi}{\lambda} |\vec{b}_{j}| \alpha_{FT,j} \right),
\end{equation}
where $\alpha_{FT}$ is the one-dimensional coordinate system in parallel to the $j$-th imaging baseline vector. The origin of the coordinate system is the center of the field of view. Comparing Equation \ref{eqn:ft_image} with Equation \ref{eqn:reconstructed_corr_final}, we found that both approaches are analytically equal if the sky consists of multiple point sources. A continous source can be reconstructed only through Fourier transform of the interferometric signal (i.e. complex visibility). However, focusing on the fact that the transmission pattern of the sky induced by the nulling baseline can be utilized to increase the signal-to-noise ratio of the planet detection, the Fourier transform method requires the condition that the imaging and nulling baselines are aligned for better planet detection. In contrast, the correlation method can be applied to any telescope configuration. Thus, the correlation method would be more utilized for planet detection, compared to the Fourier transform method \citep{Matsuo+2011}.  

\subsection{Extraction of planet spectrum (phase-space decomposition)} \label{subsec:characterization}

Once the planet light is successfully detected in the reconstructed two-dimensional image, we can estimate the planet spectrum. We calculate how the planet light is modulated while rotating the baseline based on the two-dimensional planet positional information. The planet signal for each spectral channel can be extracted from the data collected during baseline rotation. However, because the long-term fluctuation of OPD correlates with the modulation of the planet signal during baseline rotation, the reconstruction of the plaent spectrum is more easily affected by a long-term OPD error. In other words, the characterization of the planet light is more challenging than planet detection. Therefore, if the stellar leak is much brighter than planet signal, the bright stellar leak has to be subtracted before extracting the planet signal from the data. 

Here, as introduced in Section \ref{subsec:overview}, how the stellar leak changes while rotating the baseline could be measured from the data at short wavelengths. This is because the signals of warm and temperate planets except for a hot Jupiter are negligible compared to the stellar leak at short wavelengths. If we confirm from the reconstructed image that the stellar leak mainly contributes to the data at short wavelengths, the wavelength dependence of the stellar leak can be modeled in the same wavelength range. After the stellar leak model is extrapolated to the longer wavelength range, the stellar leak is subtracted from the demodulated signal shown in Equation \ref{eqn:demodulated_signal}.

There are mainly two systematic error terms, the first-order phase error and the cross-term of phase and amplitude, in the demodulated signal. Because the demodulated signal of the stellar leak in Equation \ref{eqn:demodulated_signal} is characterized by $\sin^{2}\left(\frac{\pi}{\lambda} \vec{b} \cdot \vec{\theta} + \delta l_{n} \right) \sin \left( \frac{2 \pi}{\lambda} \vec{B} \cdot \vec{\theta} + \delta l_{i} \right)$, the wavelength dependence of the stellar leak could be simply expressed by a power law of wavelength under the condition that the systematic errors are much smaller than wavelength:
\begin{equation}
	\label{eqn:reference}
	I_{leak}(\lambda, t) = \left \{ a \left(\frac{\lambda}{\lambda_{0}} \right)^{\alpha} + b \left(\frac{\lambda}{\lambda_{0}} \right)^{\beta} \right \} I_{leak}(\lambda_{0}, t),
\end{equation}
where $\lambda_{0}$ is the reference wavelength, and $I_{leak}(\lambda, t)$ is the stellar leak model in the demodulated signal at the wavelength $\lambda$ as a function of time, $t$. We note, however, that different wavelength dependencies may exist in the long wavelength range because a coating dispersion error or a pupil shear error could impact at long wavelengths. Because the coating dispersion and pupil shear errors drastically decreases in the longer wavelength regime, the subtraction of the estimated stellar leak model from the demodulated signal would less impact the reconstructed planet spectrum at long wavelengths. We also note that thanks to the bright stellar leak at short wavelengths, the systematic aberration could be modeled from the intensity and wavelength dependence of the stellar leak, which is a similar work as measuring coronagraphic low-order aberrations \citep{Guyon+2009}.

After subtracting the stellar leak from the demodulated signal shown in Equation \ref{eqn:demodulated_signal}, the planet spectra are reconstructed from the residual data through the SVD method. The matrix equation for the $i$-th spectral element can be written as
\begin{equation}
	\label{eqn:matrix}
	O_{i} = R_{i} I_{i},
\end{equation}
where $O$ is the vector composed of the observed data, $R$ is the matrix of the response function of objects (i.e. sky transmission of object), and $I$ is the vector of the input sky. In order to reconstruct the spectra of the planets, we solve the matrix equation for each spectral channel.

After the observed data is subtracted from the averaged value over the baseline rotation, the $O$ vector of the $i$-th spectral element is 
\begin{equation}
	\label{eqn:reconstruct_spectrum}
O_{i} = \left(
\begin{array}{cccc}
O_{i,1}  \\
O_{i,2} \\
\vdots \\
O_{i,N_{j}}
\end{array}
\right),
\end{equation}
where $O_{i,j}$ is the observation data of the $j$-th azimuth angle for the $i$-th spectral element. The number of the elements for the $O$ vector is $N_{j}$, corresponding to the number of the collected data during baseline rotation. The response matrix for the $i$-th spectral element, $R_{i}$, is written as
\begin{equation}
R_{i}  = \left(
\begin{array}{ccc}
R_{1,1} & \ldots & R_{1, N_{p}}  \\
\vdots &  & \vdots  \\
R_{N_{j},1} & \ldots & R_{N_{j},N_{p}}
\end{array}
\right),
\end{equation}
where we assumed that the continuum component is removed from the observed vector, $O_{i}$, by the subtraction of the two chop states. The $R$ matrix is a $N_{j} \times N_{p}$ matrix. Each component is as follows:
\begin{eqnarray}
R_{1,1} &=& \sin^{2}\left( \frac{\pi}{\lambda_{i}} \vec{b}_{1}\cdot \vec{\theta}_{p,1} \right) \sin\left( \frac{2\pi}{\lambda_{i}} \vec{B}_{1}\cdot \vec{\theta}_{p,1} \right) \\ \nonumber
R_{1,N_{p}} &=& \sin^{2}\left( \frac{\pi}{\lambda_{i}} \vec{b}_{1}\cdot \vec{\theta}_{p,N_{p}} \right) \sin\left( \frac{2\pi}{\lambda_{i}} \vec{B}_{1}\cdot \vec{\theta}_{p,N_{p}} \right) \\ \nonumber
R_{N_{j},1} &=& \sin^{2}\left( \frac{\pi}{\lambda_{i}} \vec{b}_{N_{j}}\cdot \vec{\theta}_{p,1} \right) \sin\left( \frac{2\pi}{\lambda_{i}} \vec{B}_{N_{j}}\cdot \vec{\theta}_{p,1} \right) 
\end{eqnarray}
The vector of the input sky for the $i$-th spectral element is 
\begin{equation}
	\label{eqn:reconstruct_spectrum_2}
I_{i} = \left(
\begin{array}{cccc}
I_{p,1}(\lambda_{i}, \vec{\theta}_{p,1}) \\
I_{p,2}(\lambda_{i}, \vec{\theta}_{p,2}) \\
\vdots \\
I_{p,N_{p}}(\lambda_{i}, \vec{\theta}_{p,N_{p}})
\end{array}
\right).
\end{equation}
The number of elements for the $I$ vector is $N_{p}$. When the number of the observed data is much larger than that of the elements for the input matrix, the $N_{p}$ planet signals for each spectral element can be decomposed by the SVD method. 

Finally, we need to emphasize that there are several ways to decompose the planet signals and unknown stellar leaks under the condition that the planets are successfully detected. This decomposition could be solved using modified orthogonal projections, or kernels, such as used in \citep{Romain+2020}. They preserve important properties of the covariance matrix of errors, and this decomposition therefore is well suited for further data whitening approaches. 

We could also combine PSSD with wavelet-based signal reconstruction methods \citep[e.g.,][]{delSer+2018}, either by suppressing wavelet coefficients at different levels, corresponding to unwanted signals, or convolving them with specially design convolution kernels. It is expected that while the low-frequency signal is mainly caused by the systematic OPD residual, the mid- to high-frequency signals are caused by off-axis point sources and stochastic noises. Knowing the models of the systematic errors and stochastic noises makes it possible to suppress sections of the wavelet decomposition connected to these signals and then, using the inverse wavelet transform, rebuild the planet signal. Reconstruction of the signal will depend on the cadence and its level with respect to noise. With advanced techniques we could expect signal recovery even its contribution is up to $\sim 10-30 \%$.

%In contrast, principal component analysis (PCA) does not require any assumptions for fitting the stellar leak in the whitening process. We note that the PCA is the special case of the SVD method. The eigenvalues of the covariance matrix simply correspond to the singular values of SVD. 

%The Angular Differential Kernel method based on PCA \citep{Romain+2020} would be another potential approach for extracting the planet light in the presence of the stellar leak. 

%The combination of more than two types of functions effectively may separate the planet spectrum from the stellar leak (see Section \ref{subsec:limitation}). 

\section{Simulations} \label{sec:simulation}
We performed numerical simulations to check the feasibility of PSSD under the LIFE baseline scenario. First, we briefly explain the simulation setup regarding the target system and instrument. Next, we show the results generated by PSSD under the ideal condition where only the astronomical noise contributes to the data as shot noise. Finally, we include a long-term systematic OPD error in the simulations and show its impact on PSSD.   

\subsection{Setup} \label{subsec:setup}
\begin{figure}
	 \centering
	\includegraphics[scale=0.1,height=5cm,clip]{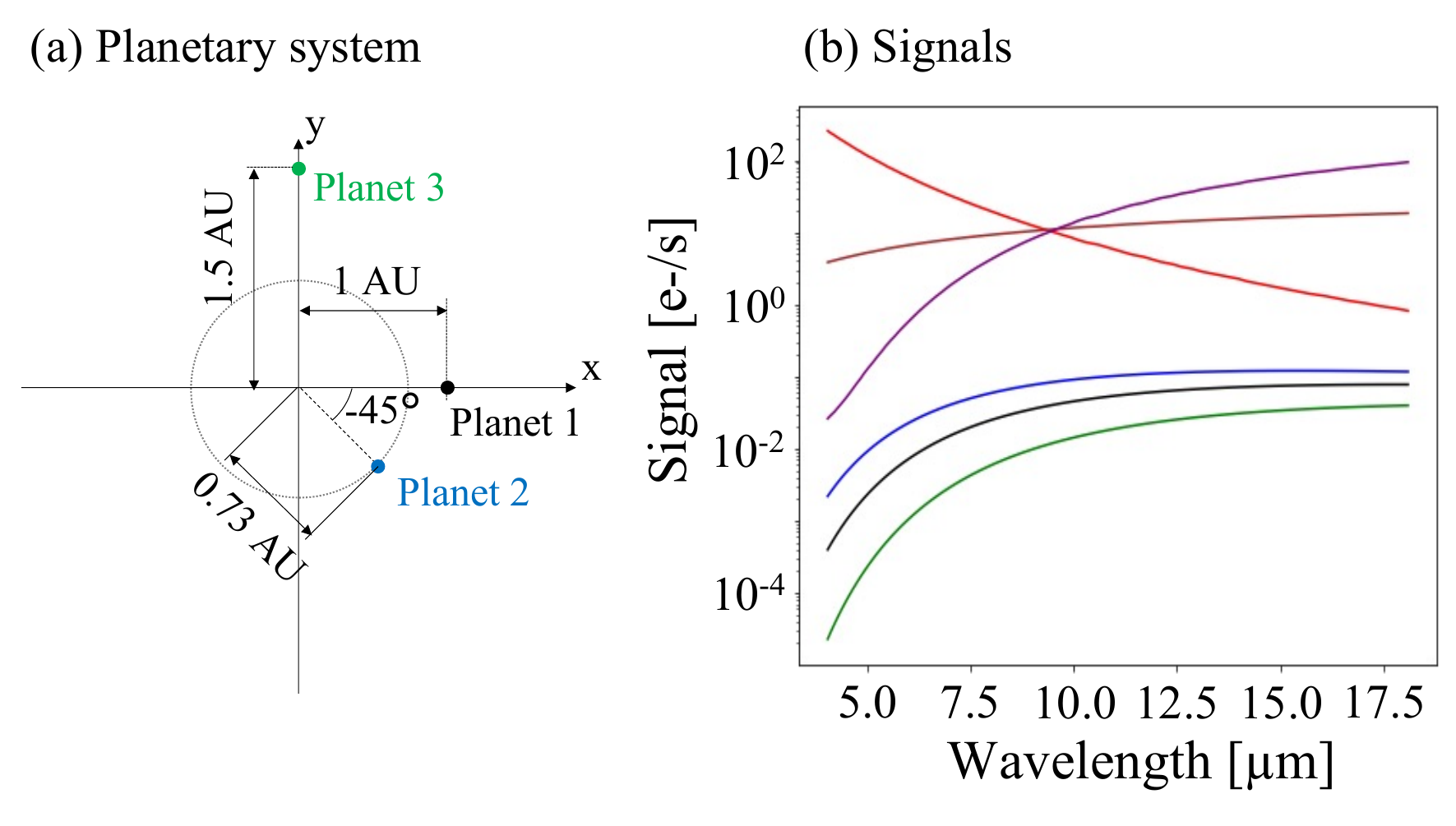}
	\caption{Target planetary system. (Left) Configuration of three planets.  (Right) Signals of planet P1 (black), planet P2 (blue), planet P3 (green), the nulled host star (red), local zodiacal light (brown), and exozodiacal light (purple) with a resolving power of 50 per a unit of time. Target system and instrument parameters are compiled in Tables \ref{tab:target_system} and \ref{tab:instrument}.}
	\label{fig:setup}
\end{figure}

The distance of the considered target is 10 $pc$. The target system consists of three Earth-sized ($R_{\oplus}$) planets, a Sun-like star with a Sun radius ($R_{\odot}$) and an effective temperature of 5778 $K$, and an exozodiacal dust disk. The semi-major axes of the three planets are 1, 0.73, and 1.5 $AU$, which are the same as those of Earth, Venus, and Mars. Given that the effective temperatures of the three planets are simply proportional to the inverse square root of the semi-major axis, the temperatures of planets P1, P2, and P3 were set to 285, 330, and 232 $K$, respectively. All of the target objects were assumed to emit blackbody radiation. The orbital phases of the three planets were set to 0, -45, and 90$^{\circ}$. The phase angle of 0$^{\circ}$ points along the positive $x$-axis, where $(x, y)$ is the coordinate system of the sky in the unit of $AU$. The arrangement of the three planets is shown in Figure \ref{fig:setup}(a). The exozodiacal light is equal to three times that of the solar system \citep{Ertel+2020}. The surface brightness of the exozodiacal light is generated based on the previous model \citep{Kennedy+2015}, which is applied to the software tool, LIFEsim \citep{Dannert+2022}. Table \ref{tab:target_system} compiles all the parameters of the target system. 

We employ a dual-Bracewell nulling interferometer with imaging and nulling baselines of 87.3 $m$ and 14.55 $m$ (see Figure \ref{fig:configuration}) so that the maximum of transmission is achieved for the center of the habitable zone around a Sun-like star at 10 $pc$ at a wavelength of 15 $\mu m$ \citep{Quanz+2022}. The diameter of each telescope is 2 $m$, and the imaging and nulling baselines are perpendicular to each other. Although the observing wavelength ranges from 4 to 18.5 $\mu m$, the same as the LIFE baseline, we limited the wavelength range to larger than 8 $\mu m$ in the planet detection phase. The reason is that the bright stellar leak is more than 100 times brighter than the light of the planets we consider at short wavelengths and deteriorates the performance of PSSD. We note that the shorter wavelengths are effective in looking for inner planets because of the combination of higher spatial resolution and brighter planets in that wavelength range. PSSD needs to optimize the wavelength range used for planet detection based on what type of planets we find. 

We set the resolving power of the spectrum to 50 for both planet detection and their characterization. The minimum resolving power for planet detection is determined by the required field of view. When the resolving power is 50, the field of view is 1.14 arcsecond at 10 $\mu m$. And for the characterization a spectral resolution of 30-50 was suggested in \cite{Konrad+2022} in order to detect the various molecules in an Earth-twin atmosphere. The total throughput was set to 0.035, given that the instrument throughput and quantum efficiency are 0.05 and 0.7, respectively. The integration time was set to 55 $h$ for planet detection and 75 days for planet characterization, respectively. 

%The lower resolving power is preferable for planet characterization in terms of the signal-to-noise ratio. However, 
%Scientifically one would expect a higher spectral resolution for the characterization phase than for the detection phase because we completely ignored the spectral resolution during the search phase as we summed the signal over the full wavelength range. We note that

We do not assume a continuous rotation but a discrete rotation in steps of one degree, where the spacecraft come to a halt before rotating again by one degree, because of computational cost. We note that the continuous rotation provides a better reconstruction thanks to the continuous U-V coverage compared to the discrete rotation. Assuming that the baseline rotates by 360$^{\circ}$ at a one-degree interval, the integration time for each baseline is 550 and 180,000 seconds for planet detection and characterization phases, respectively. In addition to the ideal observing case, we studied the feasibility of PSSD under systematic OPD error. Although there are mainly the first-order phase and the phase-amplitude cross-term in the demodulated signal \citep{Lay+2004}, only the former was considered in this simulation. We note, however, that this simulation on the feasibility of PSSD is not largely impacted by the phase-amplitude cross-term because the two systematic components have a similar frequency dependence of $\frac{1}{f}$, called "pink noise", where $f$ is the frequency. The root-mean-square (RMS) of the OPD error was set to 0.75 $nm$, corresponding to the LIFE baseline scenario for the case of only phase error \citep{Dannert+2022}. The baseline value is larger than that for the case of both phase and amplitude errors. We also consider 5, 10, and 15 times the baseline values, 3.8, 7.5, and 11.3 $nm$ RMS errors, to investigate the limitation of PSSD in Section \ref{sec:discussion}. Table \ref{tab:instrument} compiles all the instrumental parameters.

Figure \ref{fig:setup}(b) shows the astronomical signals obtained by the Bracewell nulling interferometer under the above-observed conditions. The stellar leak and background, such as the local zodiacal and exozodiacal light, cover the modulations of the three planets. When there is no OPD error, all astronomical signals contribute to the data as the shot noise. We perform the numerical simulations under the ideal condition in Section \ref{subsec:ideal_condition} and then consider the fluctuation of the stellar leak due to the systematic OPD error in Section \ref{subsec:systematic}. 

%The main reason why the OPD error is considered in this simulation is that only the OPD error moves the central star to an off-axis and leaves the stellar leak in the subtraction of the two chop states. The other instrumental noises, such as the amplitude error, polarization effect, and relative spatial shift between two apertures \citep{Lay+2004}, contribute to planet detection as the statistical noise unless they fluctuate in the timescale of the interferometric chopping. As shown in Section \ref{subsec:systematic}, although the increase of the statistical noise worsens the signal-to-noise ratio of planet detection, the residual stellar leak in the subtraction of the two chop states prevents us from measuring the planet light. In addition, even though we consider the other instrumental noises in the simulation, the residual stellar leak due to the OPD error is dominant over those of the other instrumental noises because the timescale of the interferometric chopping is much shorter than that of the baseline rotation.

%\begin{figure}
%	 \centering
%	\includegraphics[scale=0.1,height=5cm,clip]{signal.pdf}
%	\caption{Signals of planet P1 (black), planet P2 (blue), planet P3 (green), the nulled host star (red), local zodiacal light (brown), and exozodiacal light (purple) with the resolving power of 50 per a unit of time. }
%	\label{fig:signals}
%\end{figure}

\subsection{Results} \label{subsec:results}
\subsubsection{Ideal condition} \label{subsec:ideal_condition}
\begin{figure}
	 \centering
	\includegraphics[scale=0.1,height=7cm,clip]{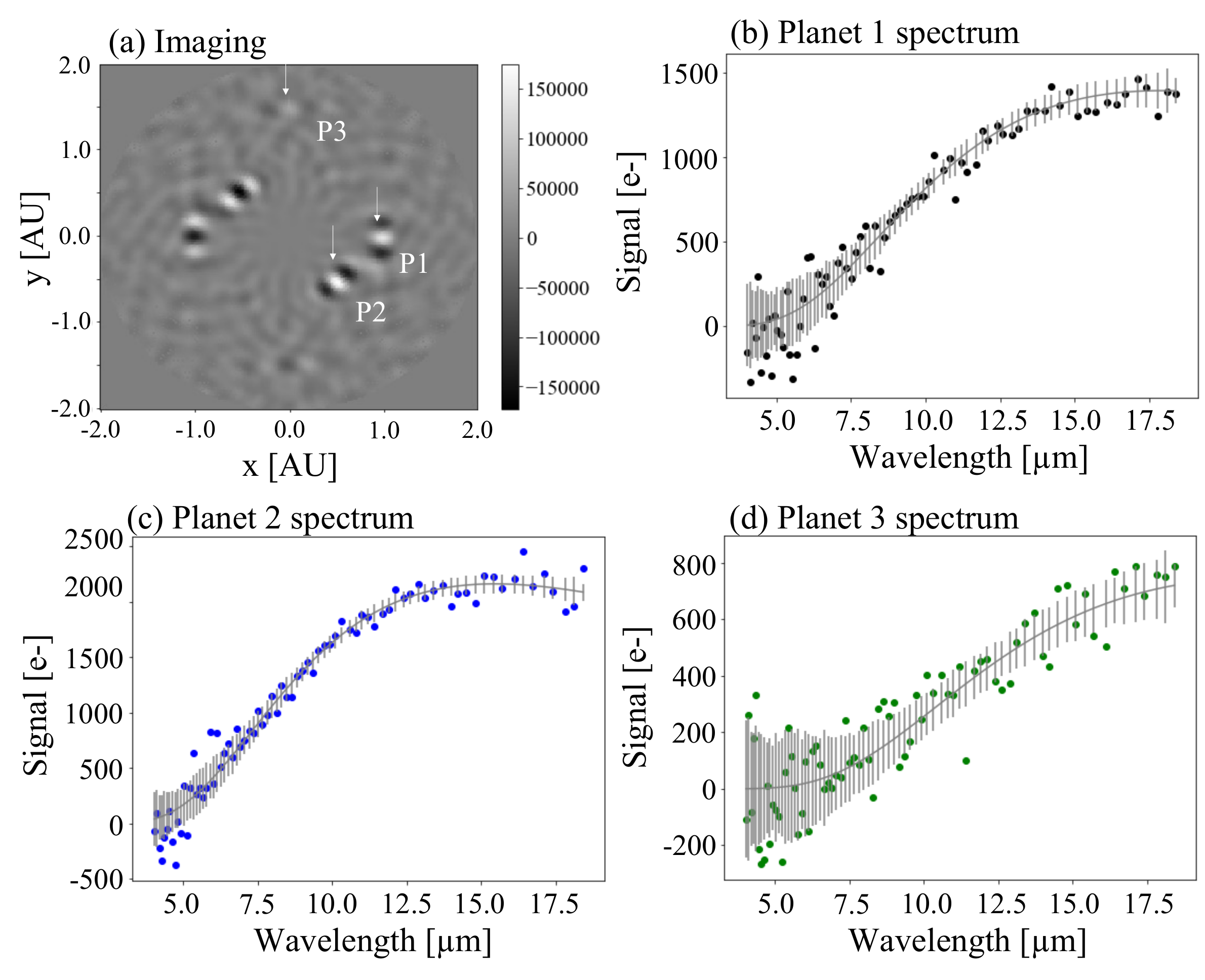}
	\caption{Image and spectrum reconstruction under photon-noise limited condition. (a) Reconstructed two-dimensional image. The white arrows denote positions of planet P1, planet P2, planet P3, respectively. The integration time was set to 55 $h$. The unit of the color bar is the number of photoelectrons. Reconstructed spectra of three planets (b) P1, (c) P2, and (d) P3 for an integration time of 75 days. The grey line and grey vertical bar of each panel show the input model and the standard deviation of each data point derived through 100 numerical simulations, respectively.}
	\label{fig:image_spectrum_reconstruction_random_error}
\end{figure}

We first show the feasibility of PSSD under the ideal condition, in which only shot noise exists due to astronomical sources. After collecting data while rotating the baseline by 360$^{\circ}$ in steps of 1 degree, we generated a two-dimensional image through processes [1] to [4]. Figure \ref{fig:image_spectrum_reconstruction_random_error}(a) shows the reconstructed two-dimensional images for an integration time of 55 $h$. The signal-to-noise ratios for the detection of planets under the ideal condition are compiled in Case 1 of Table \ref{tab:snr}. We successfully detected signals of planets P1 and P2 with signal-to-noise ratios of 10.8 and 14.6, respectively. The higher temperature of planet P2 allows us to obtain higher signal-to-noise ratio compared to that of planet P1. In contrast, the signal-to-noise ratio of planet P3 is only 3.6 because of its lower temperature. We require a longer integration time to achieve the signal-to-noise ratio of five for planet P3.

The signal-to-noise ratio is defined as the ratio of the planet signal to the starndard deviation at the same angular distance as its planet. In order to calculate the standard deviation, a two-dimensional image without the planet signals is generated and divided into annular rings. The noise floor is calculated as the standard deviation for each annular ring. We note that the absolute value of the signal-to-noise ratio cannot be directly compared with that calculated by \cite{Dannert+2022} because PSSD reconstructs both the signal and noise in a different way.

Next, we reconstructed spectra of the three planets through processes of [6], [9] and [10], assuming that the planet positions are correctly obtained. Figures \ref{fig:image_spectrum_reconstruction_random_error}(b), (c), and (d) show the reconstructed spectra of the three planets. The spectra of planet P1 and planet P2 are consistent with the input spectra (solid gray lines). We also derived for each data-point the average and standard deviation by performing the numerical simulations 100 times.  The signal-to-noise ratio of the spectrum for planet P1 agrees with that of the previous study \citep{Konrad+2022}. LIFE could detect the methane and ozone absorption bands at 7.6 and 9.6 $\mu m$ with signal-to-noise ratios of approximately 5 and 15, respectively. This combination of simultaneously detected absorption features is thought to be a good indicator of a non-equilibrium atmosphere caused by biological activity on the planet \citep{Kasting+2014}. 

On the other hand, the signal-to-noise ratio worsens at shorter wavelengths than 7.5 $\mu m$. This is because the stellar leak drastically increases due to the narrower null pattern on the sky in the shorter wavelength range. The planet signals rapidly decrease at the same time (see Figure \ref{fig:setup}(b)). In addition, PSSD also obtained the entire spectrum of planet P3 but at a signal-to-noise ratio lower than three except for wavelengths longer than 10 $\mu m$ due to its faintness compared to the other planets. Table \ref{tab:snr_characterization} compiles the signal-to-noise ratios of the reconstructed spectra at wavelengths of 5, 7.5, 10, and 15 $\mu m$.

Although planet P3 was not detected for an integration time of 55 $h$, the signal of planet P3 could be obtained while integrating the data in the characterization phase, and its position would be well determined. Planets P1 and P2 were also detected with higher signal-to-noise ratios compared to those in the planet detection phase, which can reduce the systematic errors of the reconstructed spectra due to the estimation errors of the planet positions. 

Thus, we confirmed that PSSD could detect the planet signals and characterize their atmospheres under the ideal condition, in which the data is affected only by the shot noise due to the stellar leak, background, and planets. 

\subsubsection{Systematic error} \label{subsec:systematic}

We consider instrumental noise and investigate its negative impact on planet detection and characterization. In order to evaluate it, we included an OPD error in the numerical simulations as the instrumental noise. We note, however, that the phase-amplitude cross-term also constributes to the long-term stellar fluctuation in the demodulated signal under the existence of the amplitude error \citep{Lay+2004}. Because both systematic components have the same spectrum in terms of the time domain \citep[e.g.,][]{Dannert+2022}, the phase-amplitude cross-term would not significantly impact the results. We assumed that the systematic noises have a dependency of $\frac{1}{f}$, where $f$ shows the frequency. According to \cite{Dannert+2022}, when the OPD RMS error is larger than 0.75 $nm$, the instrumental noise is dominant over the statistical noise (i.e. fundamental noise) from the astronomical objects at the shortest wavelength. Because the systematic OPD error is much smaller than the observing wavelength, the amount of stellar leak is proportional to the OPD error for each spectral element. The OPD error impacts the null depth and leaves the stellar leak in the subtraction of the two chop states (Equation \ref{eqn:demodulated_signal}). While the former contributes to the data as the Poisson noise, the latter affects the planet signal during baseline rotation, which prevents us from reconstructing the planet spectrum. In our simulations, we added the same OPD error to the imaging and nulling baselines to reduce the calculation cost. 

\begin{figure}
	 \centering
	\includegraphics[scale=0.1,height=7cm,clip]{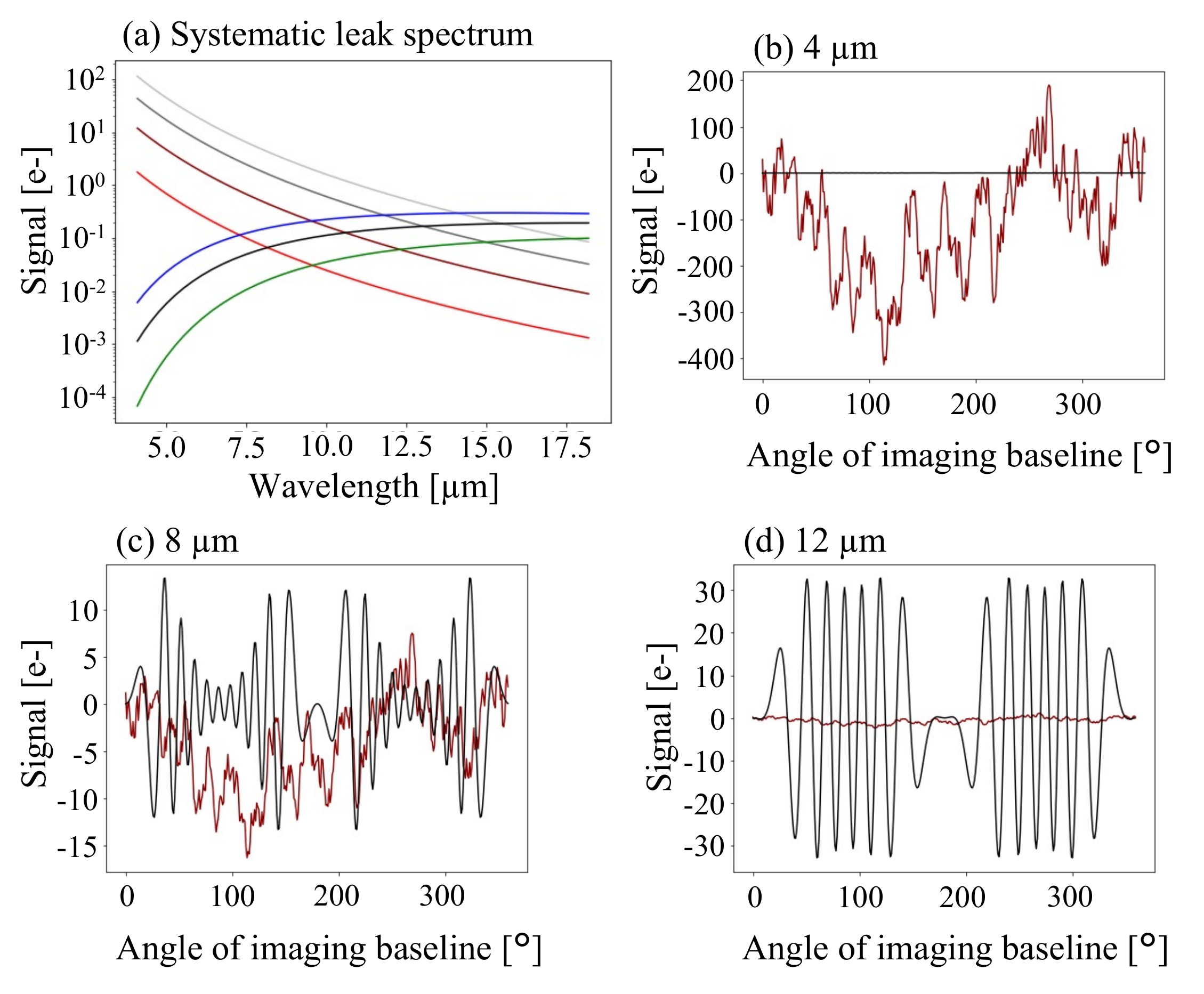}
	\caption{Systematic stellar noise. Panel (a) shows comparison of signals of the three planets P1 (black), P2 (blue), and P3 (green) with the nulled stellar leaks left in the subtraction of the two chop states over the entire wavelength range. The red, brown, gray, and light gray lines represent the stellar leaks for systematic OPD RMS errors of 0.75, 3.8, 7.5, and 11.3 $nm$. Panels (b), (c), and (d) show fluctuation of the stellar leak due to an OPD RMS error of 0.75 $nm$ (red) and the demodulated signal of planet P1 (black) at wavelengths of 4, 8, and 12 $\mu m$, respectively. The integration time of each data point is 550 $s$.}
	\label{fig:signal_systematics}
\end{figure}

Figure \ref{fig:signal_systematics}(a) compares the spectra of the three planets with the stellar leak left after subtracting the two chop states in the entire wavelength range (i.e. 4 - 18.5 $\mu m$). The stellar leak drastically increases at the shorter wavelengths because of the shallower null depth and the planet signal drops instead. It is more challenging to perform planet detection and characterization at shorter wavelengths than longer ones. Panels (b), (c), and (d) of Figure \ref{fig:signal_systematics} compare the modulated signal of planet P1 with the stellar leak at 4, 8, and 12 $\mu m$ as a function of the azimuth of the imaging baseline. Although the systematic OPD RMS error of 0.75 $nm$ does not affect the planet signal at 12 $\mu m$, the stellar leak covers the planet signal at 4 $\mu m$, which is consistent with Figure \ref{fig:signal_systematics}(a).

\begin{figure}
	 \centering
	\includegraphics[scale=0.1,height=7cm,clip]{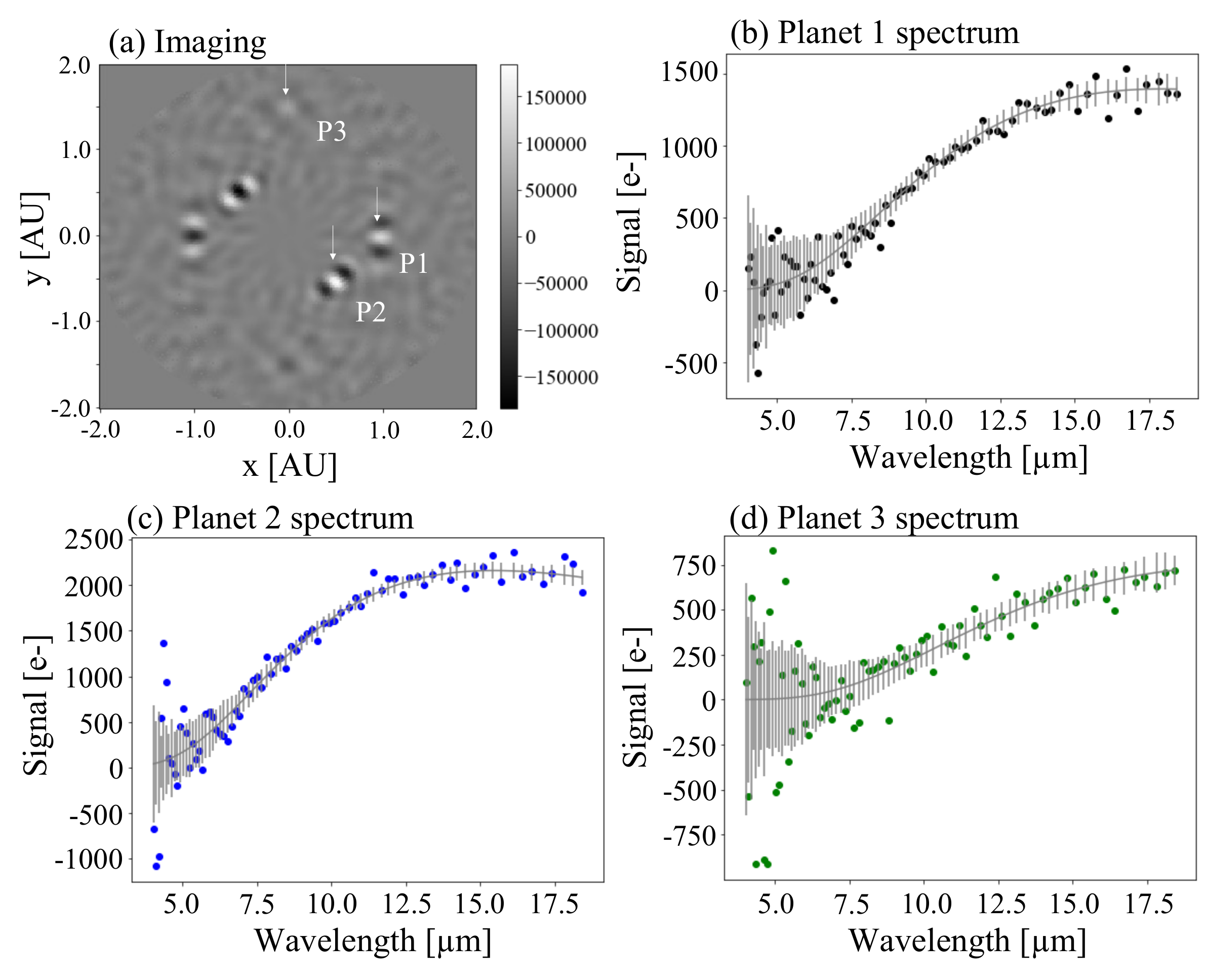}
	\caption{Same as Figure \ref{fig:image_spectrum_reconstruction_random_error} except for a systematic OPD RMS error of 0.75 $nm$.}
	\label{fig:image_spectrum_reconstruction_systematic_error}
\end{figure}

Figure \ref{fig:image_spectrum_reconstruction_systematic_error}(a) shows a reconstructed two-dimensional image under the condition that the systematic OPD RMS error is 0.75 $nm$. Thanks to robustness of PSSD against a long-term OPD error, we obtained the same signal-to-noise ratios for detecting the three planets as those for the ideal state. 

Finally, we reconstructed the planet spectra over the entire wavelength based on processes of [6], [9], and [10] without subtraction of the stellar leak from the demodulated signal in processes [7] and [8]. Panels (b), (c), and (d) of Figure \ref{fig:image_spectrum_reconstruction_systematic_error} compare the reconstructed spectra of the three planets with the input models. The reconstructed spectra are consistent with the models over the entire wavelength range, except for the shorter range. Comparing the reconstructed spectra with the ideal case, we found that the shot noise limits the performance of PSSD in the wavelength range longer than 7.5 $\mu m$. In addition, even though the systematic stellar leak is a few times brighter than the planet signals (Figure \ref{fig:signal_systematics}(a)), the signal-to-noise ratios are almost the same as those for the reconstructed spectra under the ideal condition (Case 2 of Table \ref{tab:snr_characterization}). This is because the SVD method efficiently separates the mid- to high-frequency componets induced by planets from the low-frequency component due to the systematic stellar leak. Thus, the SVD method could reconstruct the spectra of the three planets from the modulations of the planet signals during baseline rotation while efficiently separating them from the long-term systematic stellar leak.

%\begin{figure}
%	 \centering
%	\includegraphics[scale=0.1,height=4cm,clip]{fluctuation_obs_data.pdf}
%	\caption{Fluctuation of stellar leak (red) due to an OPD RMS error of 0.75 $nm$ and the modulated signal of planet P1 (black) during baseline rotation for wavelengths of 4 (left), 8 (middle), and 12 (right) $\mu m$. The integration time of each data point is 550 $s$.}
%	\label{fig:fluctuation_obs_data}
%\end{figure}

%\begin{figure}
%	 \centering
%	\includegraphics[scale=0.1,height=4cm,clip]{image_reconstruction_systematic_error.pdf}
%	\caption{Reconstructed two-dimensional image without (left) and with (right) a window function for the systematic OPD RMS error of 0.75 $nm$. The Blackman window is used as the window function. The unit of the color bar is the number of photoelectrons.}
%	\label{fig:image_reconstruction_systematic_error}
%\end{figure}

\section{Discussion} \label{sec:discussion}

We have confirmed thus far that PSSD detects planet light and extracts planet spectra under the existence of OPD fluctuation, which follows $\frac{1}{f}$. In this section we investigate how much noise amplitude PSSD can endure (Section \ref{subsec:limitation}) and compare PSSD with the previous method (Section \ref{subsec:comparison}).

\subsection{Robustness against a large OPD error} \label{subsec:limitation}

\begin{figure}
	 \centering
	\includegraphics[scale=0.1,height=7cm,clip]{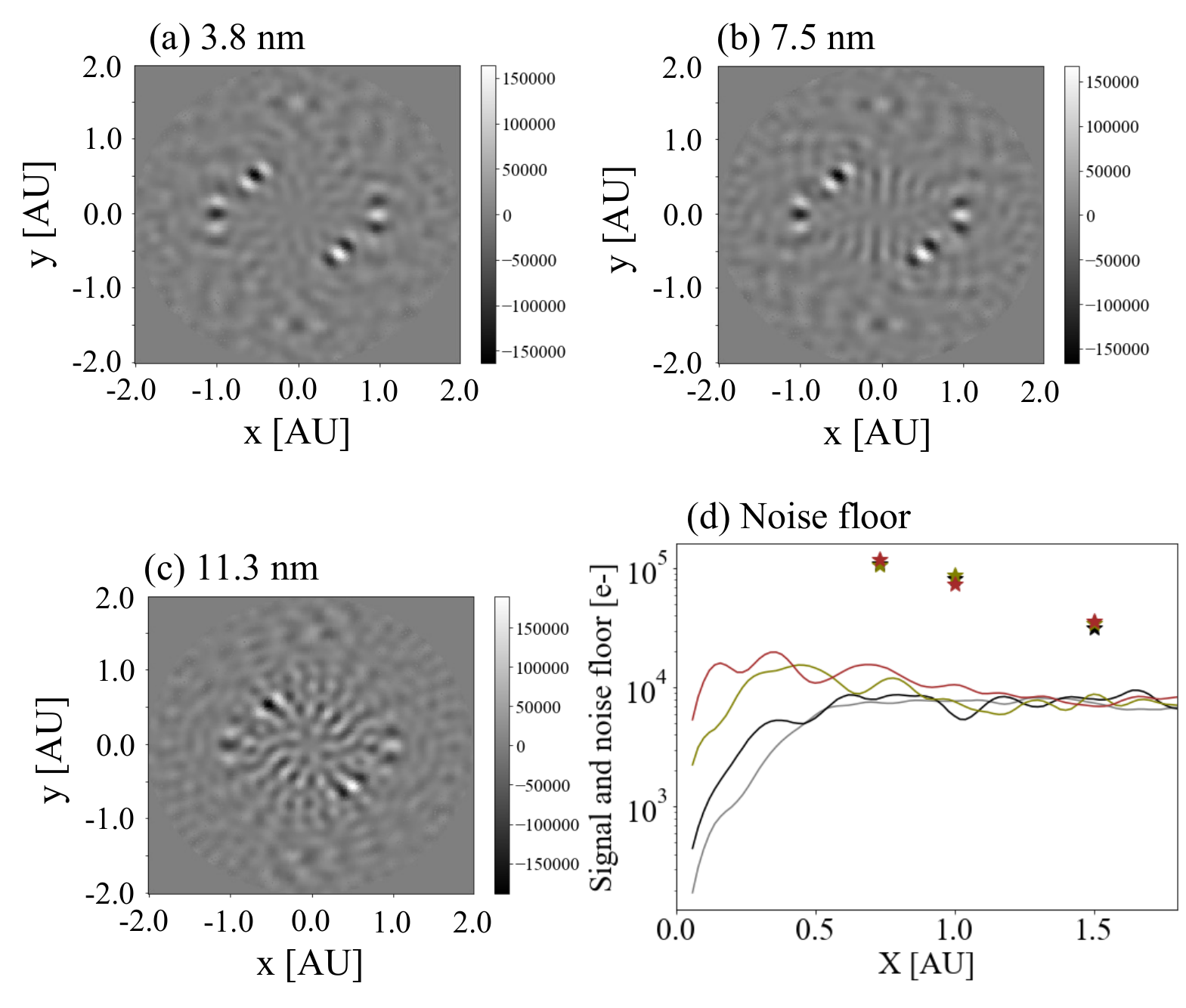}
	\caption{Image reconstruction under large systematic noises. Panels (a), (b), and (c) show reconstructed images for systematic OPD RMS errors of 3.8, 7.5, and 11.3 $nm$, respectively. Panel (d) shows the noise floors for systematic OPD RMS errors of 3.8 (black), 7.5 (yellow), and 11.3 $nm$ (brown) were compared with the planet signals (star symbol). The noise floor for each OPD error was calculated for a reconstructed two-dimensional image without the planet signals. The gray line shows the noise floor for the ideal case (i.e., only the shot noise) as a reference. Because the reconstructed planet signals are slightly affected by the OPD error, the black, yellow, and brown star symbols represent the planet signals for OPD RMS errors of 3.8, 7.5, and 11.3 $nm$, respectively.}
	\label{fig:image_reconstruction_large_systematic}
\end{figure}

We set the systematic OPD RMS error to 3.8, 7.5, and 11.3 $nm$, which are equal to the baseline values multiplied by factors of 5, 10, and 15, respectively. Figures \ref{fig:image_reconstruction_large_systematic}(b), (c), and (d) show the reconstructed images for these three different OPD errors. As the OPD error increases, systematic patterns are brighter in the centre region of each image. In contrast, the noise floor is limited by the shot noise at semi-major axes larger than 1.0 $AU$ (panel Fig. \ref{fig:image_reconstruction_large_systematic}(d)). As discussed in Section \ref{subsec:ideal_condition}, the noise floor was calculated for a reconstructed two-dimensional image without the planet signals. 

We derived the signal-to-noise ratios of the three planets for each OPD RMS error. Although the signal-to-noise ratio is gradually worsened as the OPD error increases, the signal-to-noise ratios of the planets are higher than 5 except for planet 3 (Cases 2 - 4 of Table \ref{tab:snr}). Thus, PSSD could successfully detect the inner two planets even under systematic OPD errors 15 times larger than the LIFE baseline requirement. In addition, because the shot noise limits the detection of planet 3, PSSD could also detect planet 3 with a longer integration time.

\begin{figure}
	 \centering
	\includegraphics[scale=0.1,height=10cm,clip]{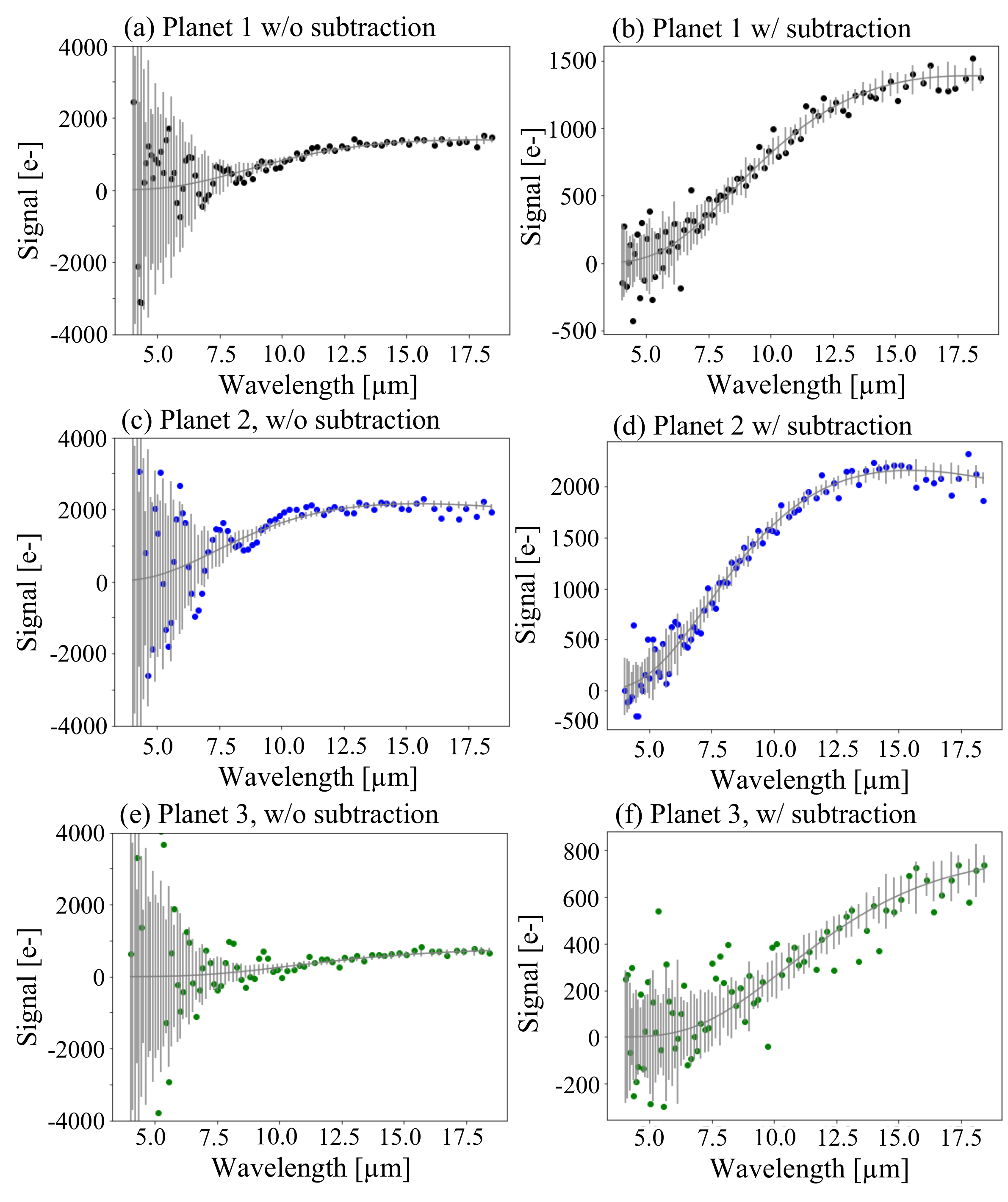}
	\caption{Spectrum reconstruction under large systematic noise. Panels (a), (c), and (e) show reconstructed spectra of planet P1, planet P2, and planet P3 only through the SVD process. Panels (b), (d), and (f) show the spectra of planet P1, planet P2, and planet P3 reconstructed through subtracting the stellar leak from the demodulated signal before the SVD method. The OPD RMS error was set to 7.5 $nm$ for both the two cases.}
	\label{fig:spectrum_reconstruction_4nm}
\end{figure}

Next, we reconstructed the spectra of the three planets for an OPD RMS error of 7.5 $nm$ by solving the matrix equation with the SVD method (process [9]; see left panels of Figure \ref{fig:spectrum_reconstruction_4nm}).  However, the OPD RMS error of 7.5 $nm$ deforms the spectra of planet P1 and planet P2 at the shorter wavelengths than 10 $\mu m$ and the spectrum of planet P2 over the entire wavelength range. This is because the planet spectra are reconstructed from the long-term modulation of the signal while rotating the data, which correlates with the systematic OPD error. Thus, characterizing the planet atmosphere is much more affected by the long-term OPD error, compared to planet detection. 

Here, we utilize the advantages of PSSD in terms of planet detection. Because a two-dimensional image can be reconstructed even for a large OPD error, we can investigate what kind of objects orbits the host star. Once we confirm that the stellar leak is dominant over the planet signals at short wavelengths, we could measure the long-term fluctuation of the stellar leak from the data at short wavelengths (process [7]) and subtract the stellar leak from the demodulated signal (process [8]). There are several steps to reconstruct the planet spectrum. First, we apply a low pass filter to the demodulated signal to decrease the impact of statistical noise on the data. Second, the wavelength dependence of the stellar leak is estimated from a large number of the data points collected during baseline rotation. Finally, based on the estimated wavelength dependence, we extrapolate the stellar leak model to the longer wavelength range and subtract it from the demodulated signal shown in Equation \ref{eqn:demodulated_signal} over the entire wavelength. The planet spectra are reconstructed through applying the subtracted data to the SVD process. Panels (b), (d), and (f) of Figures \ref{fig:spectrum_reconstruction_4nm} show the reconstructed spectra of the three planets through the above process for a large OPD RMS error of 7.5 $nm$, corresponding to ten times larger than the baseline requirement of LIFE. In this simulation, the wavelength range applied to estimation of the wavelength dependence was 4 - 5.5 $\mu m$, in which the stellar leak is more than one-hundred times larger than planet signals for an OPD RMS error of 7.5 $nm$ (Figure \ref{fig:signal_systematics}(a)). The signal-to-noise ratios for the reconstructed spectra (Case 4 of Table \ref{tab:snr_characterization}) are almost the same as those under the ideal condition (Case 1 of Table \ref{tab:snr_characterization}). Thus, if we confirm from the reconstruted image that only the stellar leak mainly contributes to the demodulated signal at short wavelengths, the planet spectra could be reconstructed.

%\begin{figure}
%	 \centering
%	\includegraphics[scale=0.1,height=7cm,clip]{fitting_residual_two_methods.pdf}
%	\caption{\textit{Left} Comparison of simulated data with a polynomial function (black) constructed from the SVD method (a) \textcolor{red}{and the wavelength-dependent fitting (c). While the raw data at 4 $\mu m$ (a) are fitted by a polynomial (see Equation \ref{eqn:matrix}), the estimated leak at 7.5 $\mu m$ was derived from the wavelength-dependent fitting.} \textit{Right} the residual between the data and \textcolor{red}{estimated leak from the two methods} (dark red) and 100 times amplified signal of planet P1 (black). The residual mainly consists of the stellar leak.}
%	\label{fig:fitting_residual}
%\end{figure}

\subsection{Comparison with previous method} \label{subsec:comparison}

We compare PSSD with the previous signal extraction through the cross-correlation or the maximum likelihood of the data obtained while rotating the baseline \citep[e.g.,][]{Angel+1997, Dannert+2022}. The main difference between the two methods is whether a one-dimensional image from each baseline is first reconstructed or a two-dimensional image is reconstructed at one time. PSSD performs the correlation of planet signal among the obtained spectrum and converts one-dimensional information into a two-dimensional image. In contrast, the previous method simultaneously finds the modulation of the planet signal in both the wavelength- and time-domains. As discussed in Section \ref{subsec:overview}, the advantages of PSSD are robustness against a large OPD error and a limited number of baselines. 

\begin{figure}
	 \centering
	\includegraphics[scale=0.1,height=7cm,clip]{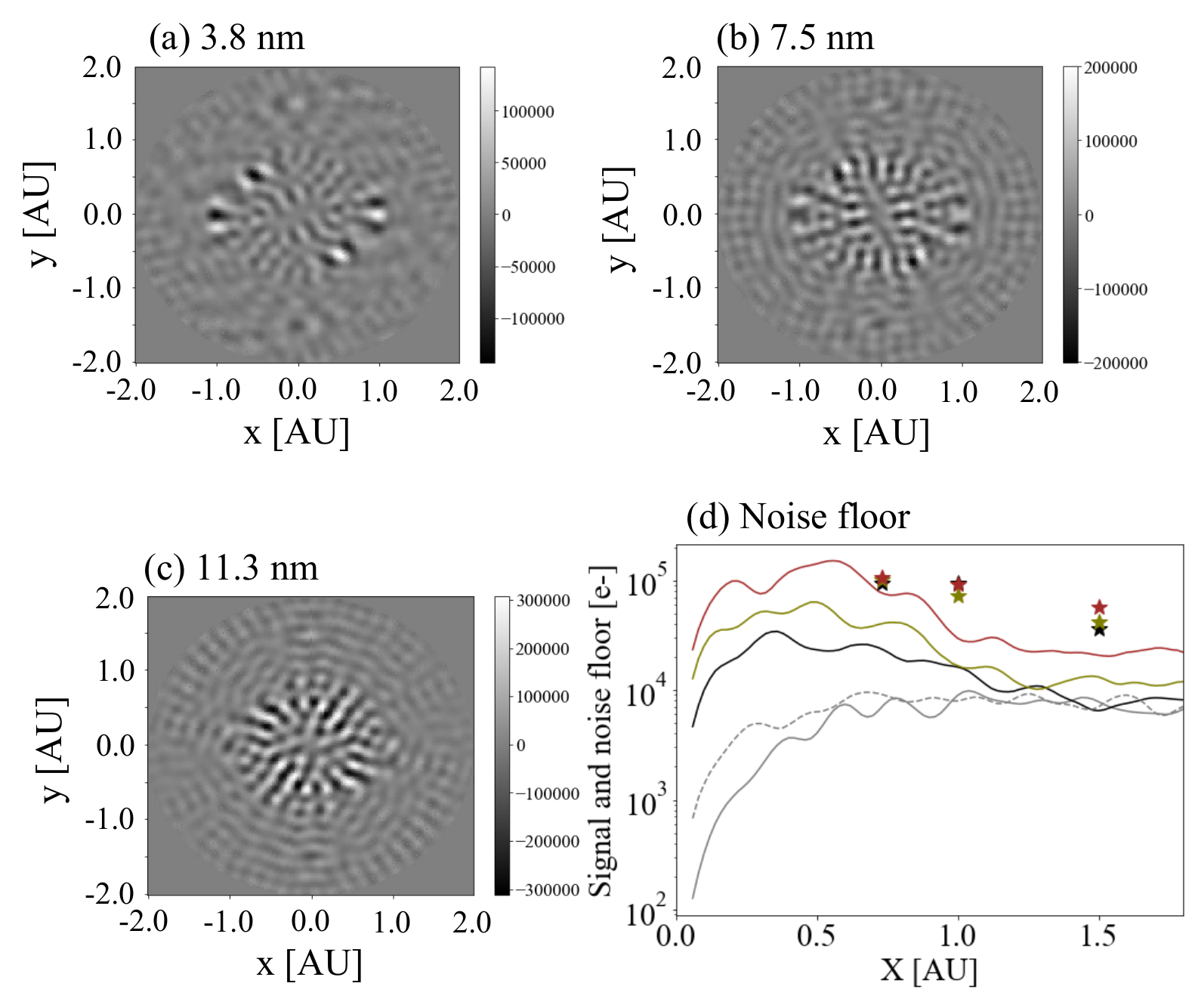}
	\caption{Same as Figure \ref{fig:image_reconstruction_large_systematic} except for using the cross-correlation method. The noise floor derived under the LIFE baseline requirement (dashed line) is added as a reference.}
	\label{fig:previous_noise_floor}
\end{figure}

Panels (a), (b), and (c) of Figure \ref{fig:previous_noise_floor} show the reconstructed two-dimensional images through the previous cross-correlation method. The images were formed based oh the following Equation:
\begin{equation}
	\label{eqn:reconstructed_corr}
	M_{corr}(\vec{\alpha}_{corr}) = \Sigma_{j}^{N_{j}} \Sigma_{i}^{N_{i}} O(\lambda_{i}) \sin \left(\frac{2\pi}{\lambda_{i}} \vec{B}_{j} \cdot \vec{\alpha}_{corr} \right) \sin^{2} \left(\frac{\pi}{\lambda_{i}} \vec{b}_{j} \cdot \vec{\alpha}_{corr} \right).
\end{equation}
Compared with Figure \ref{fig:image_reconstruction_large_systematic}, the OPD error induces brighter systematic patterns at semi-major axes smaller than 1.0 $AU$, which prevent us from detecting the two inner planets. As shown in panel (d) of Figure \ref{fig:previous_noise_floor}, the intensity of the systematic pattern is roughly proportional to the OPD RMS error in the inner region. We note that the noise floor under the baseline requirement of LIFE is almost equal to that for a systematic OPD RMS error of 0.75 $nm$, which is consistent to the previous study \citep{Dannert+2022}. The signal-to-noise ratios for the detection of planets with PSSD under an OPD RMS error of 7.5 $nm$ (Case 2 of Table \ref{tab:snr}) are almost the same as those for the cross-correlation method under the ideal case (Case 5 of Table \ref{tab:snr}). Therefore, there exists a large difference between the robustness against a large OPD error.

\begin{figure}
	 \centering
	\includegraphics[scale=0.1,height=10cm,clip]{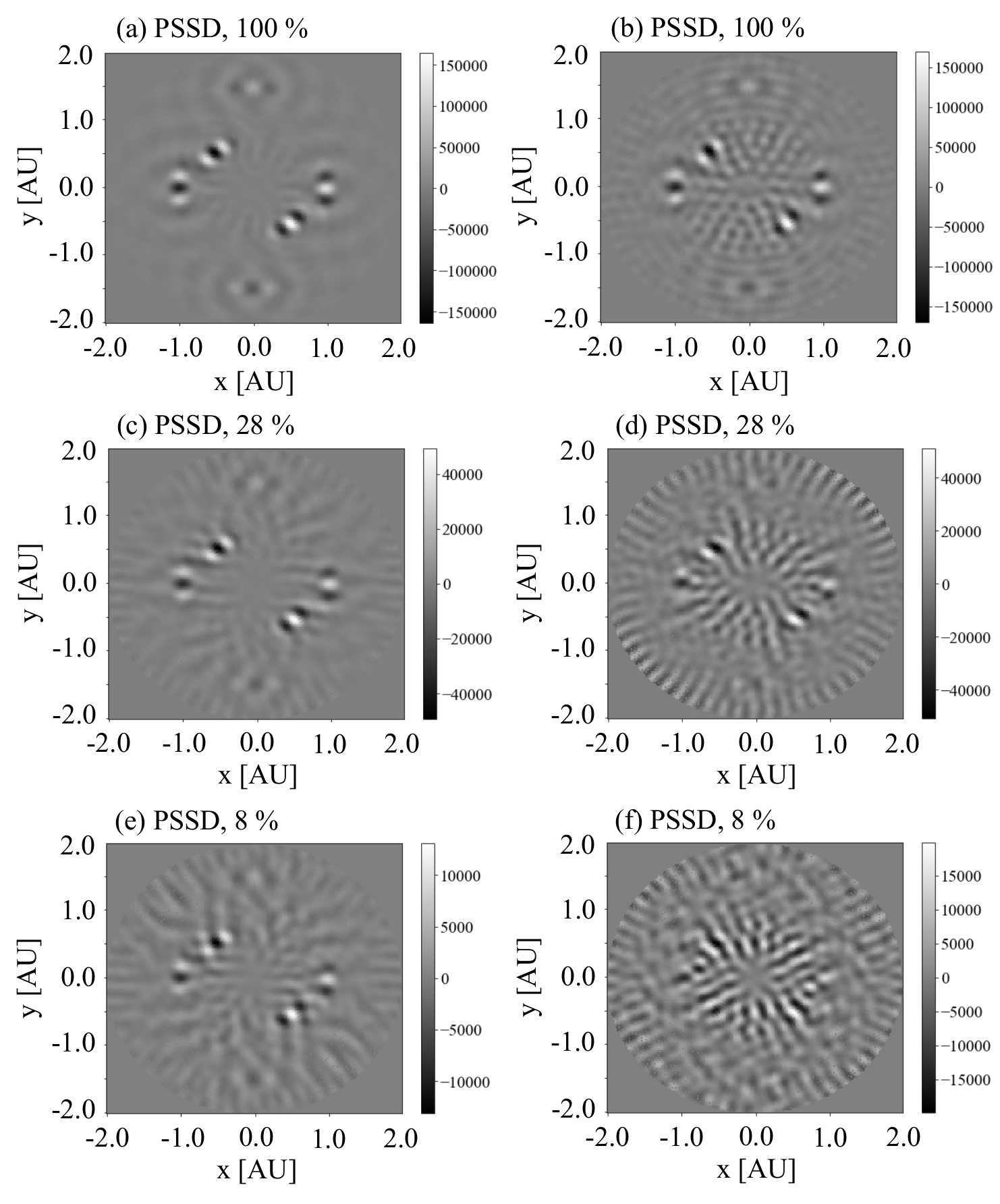}
	\caption{Images reconstructed through PSSD (left panels) and the previous method (right panels) under the condition that the number of baselines is limited by (a)(b) 100 \%, (c)(d) 28 \%, and (e)(f) 8 \%. Only a long-term systematic error was included in the simulations; shot noise was not considered while investigating how the limited number of the baselines affects the image reconstruction. The OPD RMS error was set to 3.8 $nm$.}
	\label{fig:limited_baselines_pssd}
\end{figure}

We also compare PSSD with the cross-correlation method in terms of the impact of a limited number of baselines on planet detection. The left panels of Figure \ref{fig:limited_baselines_pssd} show the PSSD reconstructed images with a limited number of baselines and under the existence of only long-term systematic error (without shot noise). In other words, the reconstructed image is not influenced by the integration time. We randomly selected available baselines, which are more sparsely distributed over 360$^{\circ}$ as the number of baselines decreases. The systematic pattern is slightly brighter as the number of baselines decreases. However, the three planets could be detected even though the fraction of the available baselines is limited to only 8 \%. In contrast, images reconstructed using the previous method were largely affected by the limited number of baselines because the modulation in the time domain was lost (the right panels of Figure \ref{fig:limited_baselines_pssd}). In addition, the artificial pattern fully covers the three planets if the number of baselines is limited to 8 \%, and the limited azimuth coverage elongates the point sources, which means that the long-term systematic error modulates the data collected while rotating the baseline.  

Thus, PSSD is much less impacted by both a large systematic error and a limited number of available baselines compared to the previous method, which could relax the requirement of LIFE. 

%Figure \ref{fig:limited_baselines_large_error} shows the reconstructed images through the two methods for the larger OPD RMS error of 7.5 $nm$, which is as two times high as those for Figures \ref{fig:limited_baselines_pssd} and \ref{fig:limited_baselines_previous}. Regarding the previous method (panels (c) and (d) in Figure \ref{fig:limited_baselines_large_error}), the planet signals are more impacted by the larger amplitude of the systematic noise. In contrast, a larger OPD RMS error is acceptable for PSSD under the limited number of baselines (Panels (a) and (b)).  

%Thus, PSSD could relax the requirements of LIFE regarding the amplitude of the OPD error and the duration of keeping the stability, which leads to a shortening of the required duration from a few days to a few minutes.

\section{Conclusion} \label{sec:conclusion}

We proposed a method for planet detection and characterization with future nulling space interferometers, such as large interferometer for exoplanets (LIFE). The proposed method is named "phase-space synthesis decomposition (PSSD)." PSSD focuses on the correlation of the planet signal over the entire wavelength range instead of that along the baseline rotation. Because a one-dimensional image parallel to the baseline can be derived for each baseline, a large number of one-dimensional images are collected after rotating the baseline. A two-dimensional image can be reconstructed by summing over the one-dimensional images. Once the two-dimensional image is obtained, a continuous equation is constructed based on the planet position information, and its solution through singular value decomposition (SVD) allows us to extract the planet spectra embedded in the stellar fluctuation. As long as the modulation of the planet signal has a different frequency from the stellar fluctuation, the SVD method efficiently decomposes the stellar leak and planet signal. PSSD provides three advantages in planet detection compared to previous methods that find a modulation of the planet signal during baseline rotation in both the wavelength- and time-domains. One is robustness against a large systematic OPD error. Because the stellar leak has a different wavelength dependence from the planet signal, only the correlation of the signal efficiently decomposes the stellar leak and the planet signal. The second is that PSSD does not correlate with a long-term fluctuation of the stellar leak because a two-dimensional image is formed by summing over one-dimensional images that require only two-phase chop states. The third advantage is robustness against a limited number of baselines.

We performed numerical simulations to investigate the feasibility of PSSD under various conditions. We put three terrestrial planets with semi-major axes of 0.73, 1, and 1.5 $AU$, corresponding to those of Venus, Earth, and Mars, respectively, around a Sun-like star at 10 $pc$. The simulation included both statistical and systematic noises. PSSD successfully detected the three planets and reconstructed their spectra for an OPD RMS error of 0.75 $nm$, which is the same as the baseline requirement of LIFE.

We also confirmed that PSSD has robustness against a large systematic OPD error. We increased the amplitude of the systematic OPD error by a factor of 5, 10, and 15. PSSD successfully detected the three planets almost without being affected by the stellar leak, even under the largest systematic OPD error of 11.3 $nm$. This is because the stellar leak is inversely proportional to approximately the fourth power of wavelength, which is largely different from the modulation of the planet signal in the wavelength domain.

In contrast, the reconstructed spectra were more affected by the long-term systematic noise than planet detection. This is because PSSD uses the modulation of the planet signal during baseline rotation to reconstruct the spectra of the three planets. The long-term systematic OPD error more easily correlates with the modulation of the planet signal. The spectra could not be accurately extracted under an OPD RMS error of 7.5 $nm$, corresponding to ten times the baseline requirement of LIFE. The signal-to-noise ratio significantly decreases in the wavelength range shorter than 7.5 $\mu m$.

Here, focusing on the fact that the planet signals are much smaller than the stellar leak at shorter wavelengths than 6 $\mu m$, we can measure the fluctuation of the stellar leak and its wavelength dependence for the data at short wavelengths. Because PSSD can successfully obtain the planet signals even under a large systematic noise, PSSD utilizes the planet position information and tells us what type of sources contribute to the signal at short wavelengths. If we confirm from the reconstructed two-dimensional image that the stellar leak is dominant over the planet signals at short wavelengths, the wavelength dependence of the stellar leak can be modeled from the data in the short wavelength range. After extrapolating the estimated stellar leak model to the longer wavelength range, the stellar leak could be subtracted from the data over the entire wavelength range. The spectra of the three planets were successfully reconstructed by applying the subtracted data to the SVD process. The signal-to-noise ratios were almost the same as those for the ideal condition.

%\textcolor{red}{We also discussed the impact of a hot Jupiter on PSSD. Because the signal of a hot Jupiter exceeds the stellar leak even at the shortest wavelengths, the stellar leak cannot be estimated from the data in the short wavelength range. Here, if both the planet position and spectrum are measured, the modulation of the planet signal while rotating the baseline can be reconstructed. The hot Jupiter can be spatially resolved from the host star thanks to the long imaging baseline. The spectrum of the planet is also derived through the SVD process based on the planet position information. We can subtract the modulation of the planet signal from the data and estimate the stellar leak for the subtracted data. The spectra of three fainter planets could be successfully obtained through the second SVD process. Thus, the impact of the hot Jupiter on PSSD could be mitigated. We note, however, that the signal-to-noise ratios of three planets decreased because the reconstructed spectrum of the hot Jupiter was not precisely reconstructed.}

Finally, we compared PSSD with a previous method that reconstructs a two-dimensional image by simultaneously fitting the modulation of the planet in the time- and wavelength domains after the baseline is rotated. Because the long-term noise is correlated with the planet signal in the time domain, systematic patterns were formed in the reconstructed image and covered the planet signals under systematic OPD errors larger than 3.8 $nm$. The signal-to-noise ratio for planet detection significantly decreased for large OPD errors compared to PSSD. This is because the noise floor increased in the inner region due to the systematic OPD error. In addition, limited azimuth coverage in the U-V plane impacted planet detection because the modulation in the time domain was lost. In contrast, PSSD can reconstruct a two-dimensional image from fewer baselines. Even in the case where the azimuth coverage of the baseline is limited to 8 $\%$, the three planets could be discovered by PSSD.

Thus, PSSD is more robust against a large OPD error and a limited number of baselines, which could relax the requirements of LIFE regarding the OPD error and the stability duration. However, this numerical simulation was performed as the first step under an ideal case that makes detection of terrestrial planets easier. We will investigate various impacts not considered in this study, such as asymmetric exozodiacal structure \citep{Defrere+2010}, and the other systematic errors, on planet detection and characterization as the next step. %\textcolor{red}{We will also develop a more sophisticated method for mitigating the impact of hot Jupiters on the reconstruction of the planet spectrum.} 

\begin{acknowledgements}
We express our sincere gratitude to Dr. Lacour for many valuable comments and suggestions on this study. Part of this work was supported by JST FOREST Program, Grant Number JPMJFR202W. Part of this work has been carried out within the framework of the National Centre of Competence in Research PlanetS supported by the Swiss National Science Foundation under grants 51NF40$\_$182901 and 51NF40$\_$205606. FD and SPQ acknowledge the financial support of the SNSF. Part of this work was carried out within the project SCIFY, which has received funding from the European Research Council (ERC) under the European Union's Horizon 2020 research and innovation program (grant agreement CoG - 866070). A.B.K. acknowledges funding provided by University of Belgrade-Faculty of Mathematics  (the contract 451-03-47/2023-01/200104), through the grants by the Ministry of Science, Technological Development and Innovation of the Republic of Serbia.
\end{acknowledgements}


\bibliographystyle{aa} 
\bibliography{manuscript.bib}

\clearpage

\begin{table}[htb]
	\begin{center}
		\caption{Parameters of the target system.}
  		\begin{tabular}{| l | c | c | c | c |} \hline 
    		& Host star & Planet 1 & Planet 2 & Planet 3 \\ \hline \hline
    		Distance ($pc$) & 10 $pc$ & - & - & -  \\ \hline 
    		Radius & 1 $R_{\odot}$ & 1 $R_{\oplus}$ & 1 $R_{\oplus}$ & 1 $R_{\oplus}$  \\ \hline    
    		Temperature ($K$) & 5778 & 285 & 330 & 232 \\ \hline  
    		Semi-major axis ($AU$) & - & 1 & 0.73 & 1.5 \\ \hline 
    		Orbital phase ($^{\circ}$) & - & 0 & -45 & 90 \\ \hline     		     		  
  \end{tabular}
  \label{tab:target_system}
  \end{center}
\end{table}

\begin{table}[htb]
	\begin{center}
		\caption{Parameters of the LIFE instrument used for planet search and characterization.}
  		\begin{tabular}{| l | c | c |} \hline 
    		& Planet search phase & Planet characterization phase\\ \hline \hline
    		Telescope diameter ($m$) & 2 & 2 \\ \hline 
    		Configuration & Dual-Bracewell interferometer & Dual-Bracewell interferometer \\ \hline    
    		Imaging baseline ($m$) & 87.3  & 87.3 \\ \hline    
    		Nulling baseline ($m$) & 14.55 & 14.55 \\ \hline        		    		
    		Instrument throughput & 0.05 & 0.05 \\ \hline  
    		Quantum efficiency & 0.7 & 0.7 \\ \hline     		   		
    		Wavelength range ($\mu m$) & 8 - 18.5 & 4 - 18.5 \\ \hline    		
    		Resolving power & 50 & 50  \\ \hline      		   		
    		Integration time & 55 hours &  75 days \\ \hline     		
			Field of view ($\frac{\lambda}{D}$) & 1  & 1 \\ \hline  
			Systematic OPD RMS error ($nm$) & 0.75, 3.8, 7.5, 11.3  & 0.75, 3.8, 7.5, 11.3 \\ \hline  			  		
  \end{tabular}
  \tablebib{\cite{Quanz+2022, Dannert+2022, Konrad+2022}}
  \label{tab:instrument}
  \end{center}
\end{table}

\begin{table}[htb]
	\begin{center}
		\caption{Signal-to-noise ratios for planet detection with the previous method and PSSD under several OPD values.}
  		\begin{tabular}{| l || c | c | c | c | c |} \hline 
    		& Method & OPD error ($nm$) & Planet 1 & Planet 2 & Planet 3 \\ \hline \hline
    		Case 1 & PSSD & 0 (only shot noise) & 10.8 & 14.6 & 3.6   \\ \hline
    		Case 2 & PSSD & 3.8 & 10.9 & 13.3 & 3.6 \\ \hline  
    		Case 3 & PSSD & 7.5 & 10.6 & 8.4 & 2.9 \\ \hline 
    		Case 4 & PSSD & 11.3 & 6.1 & 6.9 & 3.9 \\ \hline   
    		Case 5 & cross-correlation & 0 (only shot noise) & 9.5 & 11.7 & 3.6  \\ \hline    
    		Case 6 & cross-correlation & 3.8 &  4.8 & 3.1 & 4.6  \\ \hline      		
    		Case 7 & cross-correlation & 7.5 & 3.3 & 1.5 & 2.2 \\ \hline     		
    		Case 8 & cross-correlation & 11.3 & 2.1 & 0.4 & 2.1 \\ \hline       		  		 
    		    		  		     		  
  \end{tabular}
  \label{tab:snr}
  \end{center}
\end{table}

\clearpage

\begin{table}[htb]
	\begin{center}
		\caption{Signal-to-noise ratios for reconstructed spectra of three planets.}
  		\begin{tabular}{| c || c | c | c | c | c | c |} \hline 
    		& Wavelength ($\mu m$) & OPD error ($nm$) & Subtraction of stellar leak & Planet 1 & Planet 2 & Planet 3 \\ \hline \hline
    		Case 1 & 5  & 0 (only shot noise) & No & 0.24 & 0.79 & 0.02   \\ \hline 
    		& 7.5  & 0 (only shot noise) & No & 3.2 & 9.1 & 0.76   \\ \hline
    		& 10 & 0 (only shot noise) & No & 9.5 & 21.9 & 2.0 \\ \hline  
    		& 15 & 0 (only shot noise) & No & 19.0 & 16.5 & 7.5 \\ \hline  
    		Case 2 & 5  & 0.75 & No & 0.16 & 0.43 & 0.01   \\ \hline    		 
    		& 7.5 & 0.75  & No & 3.0 & 11.1 & 0.7   \\ \hline
    		& 10 & 0.75 & No & 8.9 & 21.4 & 2.2 \\ \hline  
    		& 15 & 0.75 & No & 16.3 & 16.0 & 7.4 \\ \hline  
    		Case 3 & 5  & 7.5 & No & 0.02 & 0.05 & 0.001   \\ \hline    		 
    		& 7.5 & 7.5  & No & 0.74 & 3.2 & 0.2   \\ \hline
    		& 10 & 7.5 & No & 8.0 & 13.4 & 1.6 \\ \hline  
    		& 15 & 7.5 & No & 15.9 & 19.2 & 7.5 \\ \hline   
    		Case 4 & 5  & 7.5 & Yes & 0.18 & 0.77 & 0.01   \\ \hline    		 
    		& 7.5 & 7.5  & Yes & 3.1 & 8.0 & 0.5   \\ \hline
    		& 10 & 7.5 & Yes & 10.5 & 22.8 & 2.6 \\ \hline  
    		& 15 & 7.5 & Yes & 17.2 & 23.2 & 6.7 \\ \hline    		     		   		 
    		  		 
  \end{tabular}
  \label{tab:snr_characterization}
  \end{center}
\end{table}

\end{document}